\documentclass[aps, prb, twocolumn,amsmath,amssymb,floatfix,reprint,longbibliography]{revtex4-2}
\usepackage{graphicx}
\usepackage{physics}
\usepackage{times}
\usepackage{dcolumn}
\usepackage{hyperref} 
\hypersetup{colorlinks, allcolors=blue}

\begin{document}

\title{Topological superconductivity mediated by magnons of helical magnetic states}

\author{Kristian M{\ae}land} 
\affiliation{\mbox{Center for Quantum Spintronics, Department of Physics, Norwegian University of Science and Technology, NO-7491 Trondheim, Norway}} 
\author{Sara Abnar}
\affiliation{\mbox{Center for Quantum Spintronics, Department of Physics, Norwegian University of Science and Technology, NO-7491 Trondheim, Norway}} 
\author{Jacob Benestad}
\affiliation{\mbox{Center for Quantum Spintronics, Department of Physics, Norwegian University of Science and Technology, NO-7491 Trondheim, Norway}} 
\author{Asle Sudb{\o}}
\email[Corresponding author: ]{asle.sudbo@ntnu.no}
\affiliation{\mbox{Center for Quantum Spintronics, Department of Physics, Norwegian University of Science and Technology, NO-7491 Trondheim, Norway}}

\begin{abstract}
We recently showed that spin fluctuations of noncoplanar magnetic states can induce topological superconductivity in an adjacent normal metal [M{\ae}land \textit{et al.}, \href{https://doi.org/10.1103/PhysRevLett.130.156002}{Phys. Rev. Lett. \textbf{130}, 156002 (2023)}]. The noncolinear nature of the spins was found to be essential for this result, while the necessity of noncoplanar spins was unclear. 
In this paper we show that magnons in coplanar, noncolinear magnetic states can mediate topological superconductivity in a normal metal. Two models of the Dzyaloshinskii-Moriya interaction are studied to illustrate the need for a sufficiently complicated Hamiltonian describing the magnetic insulator. The Hamiltonian, in particular the specific form of the Dzyaloshinskii-Moriya interaction, affects the magnons and by extension the effective electron-electron interaction in the normal metal. 
Symmetry arguments are applied to complement this discussion.
We solve a linearized gap equation in the case of weak-coupling superconductivity. The result is a time-reversal-symmetric topological superconductor, as confirmed by calculating the topological invariant. In analogy with magnon-mediated superconductivity from antiferromagnets, Umklapp scattering enhances the critical temperature of superconductivity for certain Fermi momenta. 
\end{abstract}

\maketitle

\section{Introduction}

Topological superconductors are currently of enormous interest due to their possible application in fault-tolerant quantum computing \cite{TopoSCrevSato, Leijnse2012TSCrev, TopoSCandSkRev, Bernevig2013, TopoQuantumCompRevModPhys, ladd2010quantumcomp}. This class of superconductors are characterized by a fully gapped superconducting state where the gap function displays a nontrivial winding. The Kitaev chain posits the existence of Majorana bound states (MBSs) at the edges of one-dimensional (1D) topological superconducting wires \cite{Kitaev2001Oct}. A possible realization of this is in superconductor/semiconductor hybrid wires. Together with spin-orbit coupling (SOC) in the semiconductor, a magnetic field turns the conventional superconductivity into topological superconductivity (TSC) \cite{Oreg2010Oct}. Experimental signatures of the existence of MBSs at the ends of such wires have been reported \cite{MicrosoftTGP}. MBSs may also appear in the core of superconducting vortices in 2D topological superconductors \cite{TopoSCrevSato, Leijnse2012TSCrev, TopoSCandSkRev, Bernevig2013, menard2019isolated}. Spatially separated pairs of MBSs form nonlocal fermionic states that are insensitive to local sources of noise. Since MBSs are non-Abelian anyons with nontrivial particle exchange statistics, braiding of MBSs can realize fault-tolerant quantum computation \cite{TopoSCrevSato, Leijnse2012TSCrev, TopoSCandSkRev, TopoQuantumCompRevModPhys}.

One proposal to realize 2D TSC is at the interface of conventional superconductors (SCs) and chiral magnetic insulators (MIs) \cite{SkTopoSCNagaosa, SkTopoSCMajoranaChen, SkTopoSCMajoranaLossFM, SkMajoranaRex, SkTopoSCMajoranaLossAFM, SkTopoSCMajoranaDagotto, SkTopoSCgarnier, SkTopoSCMajoranaMascot, ExpSkHeterostructure, TopoSCandSkRev}. An exchange interaction across the interface, between noncolinear spins in the MI and spins of electrons in the SC can be shown to effectively act like SOC and a magnetic field, facilitating the formation of TSC \cite{TopoSCandSkRev}. Still, it is found that noncoplanar spins are needed in the MI to yield fully gapped strong TSC \cite{SkTopoSCNagaosa, TopoSCrevSato}. Coplanar, helical states give nodal gaps \cite{SkTopoSCNagaosa, SkTopoSCMajoranaChen}. These nodes can be shown to be topologically nontrivial giving rise to weak TSC \cite{TopoSCrevSato, SkTopoSCNagaosa, SkTopoSCMajoranaChen}. This paper focuses on fully gapped strong TSC.

Many studies have considered replacing phonons with magnons as the paring mechanism in the search for unconventional superconductivity \cite{KargarianFMTI, HugdalTIFMAFM,  ArneFMNM,ArneAFMNM_Umklapp,EirikNMAFM, EirikTIAFM, EirikEliashberg, brekke2023interfacial, Brekke2023Aug, Maeland2023PRL, ExpMagnonInducedHeterostruct}.
In Ref.~\cite{Maeland2023PRL}, we showed that magnons from a skyrmion crystal can induce TSC in a normal metal (NM). Skyrmions are noncoplanar spin textures which are topologically protected \cite{nagaosaRev, KlauiRev2016}. Due to a nonzero winding number of the spin a large energy barrier has to be traversed to deform the skyrmion into topologically trivial magnetic states. This topological protection is cast into doubt for small skyrmions containing few spins, so-called quantum skyrmions, where the quantum nature of the spins is important. Nevertheless, it is found that quantum fluctuations stabilize skyrmions \cite{QSkOP, Roldan-Molina2015Dec}. Additionally, it has been shown that quantum skyrmions display topological magnons \cite{Roldan-Molina2016Apr, Diaz2020TopoMagnon, QSkQTPT}. A natural question then arises if the real space topological nature of skyrmions, or the topological nature of the magnons are crucial for the TSC found in Ref.~\cite{Maeland2023PRL}.

In this paper we answer both questions to the contrary by studying helical states, i.e., coplanar, noncolinear magnetic states. 
We show that a coplanar, noncolinear magnetic state along with the presence of the Dzyaloshinskii-Moriya interaction (DMI) can induce TSC in a normal metal. 
DMI originates with SOC \cite{Moriya}, which is often a component in topological superconductors. In MI/SC interfaces, the classical coupling between spins in the MI and spins in the SC can yield TSC in the case of noncoplanar magnetic states \cite{SkTopoSCNagaosa}. In our case, the superconductivity is induced by the spin fluctuations which are affected by SOC in the form of DMI. Hence, the precise structure of the magnetic ground state appears less important, as long as it is noncolinear. The noncolinear nature of the spin state facilitates Cooper pairs of electrons with the same spin, an essential ingredient in TSC \cite{BenestadMaster, AbnarMaster, Maeland2023PRL}. 
By considering two models of DMI and comparing them in terms of symmetry, we gain further insight into the origins of magnon-mediated TSC.

We consider weak-coupling superconductivity using generalized Bardeen-Cooper-Schrieffer (BCS) theory \cite{BCS, Sigrist}. The structure of the gap function is obtained by solving the linearized gap equation. We show that a time-reversal-symmetric 2D TSC state is realized in the NM by computing the topological invariant. 
The topologically nontrivial superconducting state is a result of the form of the effective electron-electron interaction mediated by the magnons of the MI.

\begin{figure*}
    \centering
    \includegraphics[width = 0.9\linewidth]{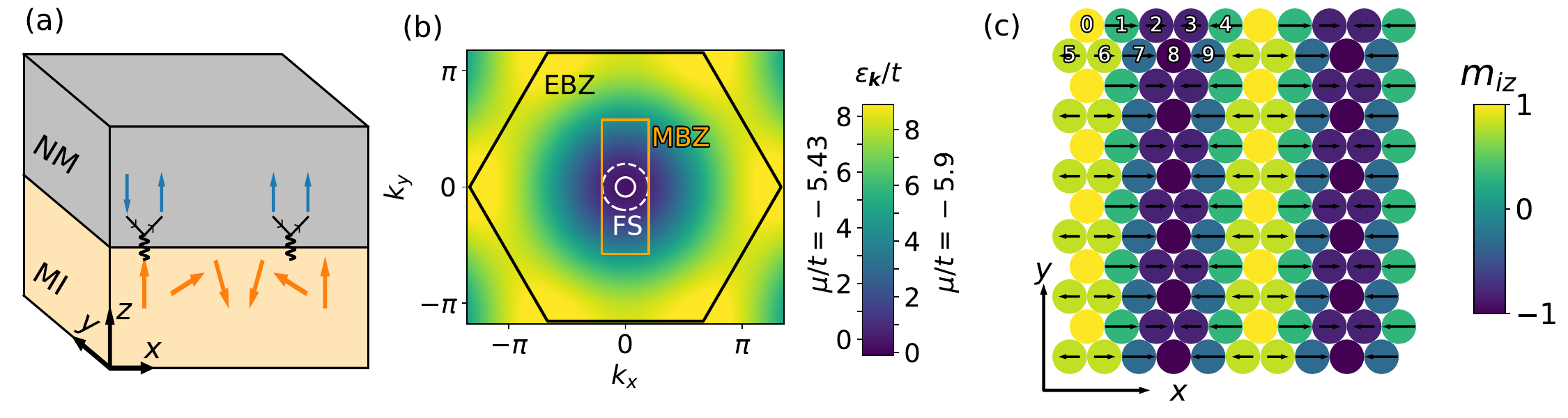}
    \caption{(a) A heterostructure of a normal metal (NM) and a magnetic insulator (MI). Orange arrows illustrate a noncolinear ground state (GS) in the MI. The interfacial exchange interaction affects electron spins (blue arrows) in the NM. (b) The NM dispersion relation along with the electron Brillouin zone (EBZ), the magnetic Brillouin zone (MBZ), and two Fermi surfaces (FSs) in white for $\mu/t = -5.9$ (solid) and $\mu/t = -5.43$ (dashed) where the FS touches the edge of the MBZ. (c) The classical GS in the MI at zero easy-axis anisotropy. Arrows and colors indicate the $x$ and $z$ components of spins, respectively. Periodic boundary conditions apply and the numbers 0 to 9 indicate the chosen sublattice numbering.}
    \label{fig:setupBZGS}
\end{figure*}

\section{Model}
We consider a heterostructure of an MI and an NM with an interfacial exchange interaction between spins in the MI and spins in the NM,
\begin{equation}
    H = H_{\text{NM}} + H_{\text{MI}} + H_{\text{int}}.
\end{equation}
The following subsections explore the components of the Hamiltonian. Figure \ref{fig:setupBZGS}(a) illustrates the heterostructure. We focus on the physics at the interface, and only consider a 2D monolayer of each material. We assume lattice matching with a triangular lattice at the interface, lying in the $xy$ plane.

\subsection{Normal metal}
We model the normal metal using a nearest-neighbor hopping $t$ and a chemical potential $\mu$, which we consider to be tunable, e.g., by gate voltages, 
\begin{equation}
    H_{\text{NM}} = -t \sum_{\langle ij\rangle \sigma} c_{i\sigma}^\dagger c_{j\sigma} - \mu \sum_{i\sigma} c_{i\sigma}^\dagger c_{i\sigma},
\end{equation}
where $c_{i\sigma}^\dagger$ creates an electron with spin $\sigma$ at lattice site $i$, located at $\boldsymbol{r}_i$. The notation $\langle ij \rangle$ indicates that sites $i$ and $j$ are nearest neighbors. The Fourier transform (FT) of electron operators is defined by $c_{i\sigma} = \frac{1}{\sqrt{N}} \sum_{\boldsymbol{k}\in \text{EBZ}} c_{\boldsymbol{k}\sigma} e^{i\boldsymbol{k}\cdot\boldsymbol{r}_i}$. The sum over (quasi)momenta $\boldsymbol{k}$ covers the first Brillouin zone of the triangular lattice. This is referred to as the electron Brillouin zone (EBZ). Furthermore, $N$ is the total number of lattice sites at the interface. The FT yields $H_{\text{NM}} = \sum_{\boldsymbol{k}\in \text{EBZ}, \sigma} \epsilon_{\boldsymbol{k}} c_{\boldsymbol{k}\sigma}^\dagger c_{\boldsymbol{k}\sigma}$, where $\epsilon_{\boldsymbol{k}} =  -\mu - 2t[\cos k_x +2 \cos(k_x/2)\cos (\sqrt{3}k_y/2)]$.

\subsection{Magnetic insulator} \label{sec:model}

In Ref.~\cite{QSkOP}, we noted the importance of the four-spin interaction to yield skyrmion states. Here, we treat an MI where we assume the four-spin interaction is negligible. Then, with no magnetic field, we expect to find helical states, also referred to as spiral states. The Hamiltonian is
\begin{equation}
\label{eq:H}
    H_{\text{MI}} = H_{\text{ex}} + H_{\text{DM}} + H_{\text{A}},
\end{equation}
where
\begin{equation}
    H_{\text{ex}}  = -J\sum_{\langle ij \rangle} \boldsymbol{S}_{i}\cdot \boldsymbol{S}_{j},
\end{equation}
\begin{equation}
    H_{\text{DM}} = \sum_{\langle ij \rangle} \boldsymbol{D}_{ij} \cdot (\boldsymbol{S}_i \cross \boldsymbol{S}_j),
\end{equation}
\begin{equation}
    H_{\text{A}} = - K\sum_i S_{iz}^2.
\end{equation}
The spin operator $\boldsymbol{S}_i$ for lattice site $i$, has magnitude $S$. We consider a ferromagnetic exchange interaction of strength $J>0$. In the DMI term, $\boldsymbol{D}_{ij} = D \hat{r}_{ij} \cross \hat{z}$, where $\hat{r}_{ij}$ is a unit vector from site $i$ to site $j$. $H_{\text{A}}$ represents an easy-axis anisotropy in the $z$ direction. Units are chosen such that Planck's constant $\hbar = 1$ and the lattice constant $a =1$.

The DMI term promotes magnetic states where the spin rotates from site to site. In competition with the exchange interaction this yields helical magnetic states. 
DMI originates with SOC, and is present in chiral magnets which break inversion symmetry \cite{Muhlbauer1stSkExp, nagaosaRev} and in magnetic monolayers on top of heavy metals \cite{HeinzeSkX, KlauiRev2016}.
Helical states have been observed in chiral magnets \cite{Muhlbauer1stSkExp}, in ultrathin films grown on heavy metals \cite{Ferriani2008Jul, Bruning2022Jun}, in van der Waals materials \cite{Meijer2020Spiral2DvdW}, and in synthetic antiferromagnets \cite{Leveille2021Spiral2DAFM}. 

For comparison, we also consider a model where the DMI vector $\boldsymbol{D}_{ij}$ is nonzero only if $\boldsymbol{r}_i$ and $\boldsymbol{r}_j$ have the same $y$ component. That is, DMI only acts on links in the $x$-direction. We refer to this as 1D DMI, and note that it could possibly arise from a special linear combination of Rashba and Dresselhaus SOC \cite{ThingstadMaster}. The case of equal DMI vector magnitude on all links is referred to as full DMI for clarity, and could arise from pure Rashba SOC \cite{ThingstadMaster}.
With 1D DMI and zero easy-axis anisotropy it is possible to find an analytic expression for the helical periodicity given $D/J$ \cite{BenestadMaster, AbnarMaster}. For full DMI, and/or nonzero easy-axis anisotropy, the magnetic ground state (GS) must be found numerically by minimizing the classical version of the Hamiltonian, i.e., by setting $\boldsymbol{S}_i = S\boldsymbol{m}_i$, where $\boldsymbol{m}_i$ is a unit vector giving the classical spin direction.

As in Ref.~\cite{QSkOP}, we tune parameters to get a preferred periodicity using trial states. Then, we use self-consistent iteration \cite{dosSantosPRB} to obtain the lowest energy state with that periodicity. Additionally, we use simulated annealing \cite{simulatedannealing} and iteration on various larger lattice sizes to search for lower energy states. If none are found, we conclude that we have found the magnetic GS, or at least a metastable state. For 1D DMI we find that $D/J = 4.98$ leads to a helical state in the $xz$ plane with a period of $\lambda_x = 5$ lattice sites in the $x$ direction, illustrated in Fig.~\ref{fig:setupBZGS}(c). For full DMI, $D/J = 2.16$ leads to a helical state in the $xz$ plane with the same periodicity. With full DMI it is less obvious that the helical state should choose the $xz$ plane. We have not found any helical states in other planes with lower energy than those in the $xz$ plane and in planes rotated by integer multiples of $2\pi/6$, where the GS degeneracy reflects the sixfold symmetry of the lattice and full DMI Hamiltonian.
By comparing energies per site of trial states, we found that the preferred periodicity is in the range $4.99 < \lambda_x < 5.01$ in both cases. The easy-axis anisotropy does not appear to affect the periodicity of the helical state, unless $K$ is the dominant energy in the Hamiltonian. Its effect is to slightly increase the magnitude of the $z$ component of all the spins, meaning that new GSs must be derived at each chosen value of $K/J$. All GSs in this paper have zero net magnetization.

The helical GSs we consider have ten unique spin directions, giving ten sublattices. Each sublattice is a rectangular lattice with primitive vectors $\boldsymbol{a}_1 = (5, 0)$ and $\boldsymbol{a}_2 = (0, \sqrt{3})$. Hence, all sublattices have the same first Brillouin zone, which we refer to as the magnetic Brillouin zone (MBZ). In reciprocal space the primitive vectors are
$\boldsymbol{b}_1 =  (2\pi/5, 0)$ and  $\boldsymbol{b}_2 = (0, 2\pi/\sqrt{3})$. Hence, the MBZ is a rectangle, as shown in Fig.~\ref{fig:setupBZGS}(b). The high symmetry points are $\boldsymbol{\Gamma} = (0,0),$ $\boldsymbol{X} = (\pi/5, 0),$ $\boldsymbol{S} = (\pi/5, \pi/\sqrt{3}),$ and $\boldsymbol{Y} = (0, \pi/\sqrt{3})$. 

In the experiment of Ref.~\cite{Leveille2021Spiral2DAFM}, the helical state periodicity was found to be approximately $190$~nm. With $a$ typically a few tenths of a nm, this is a much larger periodicity than that considered in this paper. This is because, since SOC is a relativistic effect, one usually has $D/J \ll 1$. In ultrathin films, experiments have found helical states with much shorter periodicities, namely $2.2$~nm \cite{Ferriani2008Jul, Bruning2022Jun}. From first-principles calculations, this is attributed to a relatively strong DMI and frustrations in the exchange interaction. 
While there is a strong ferromagnetic exchange interaction between nearest neighbors, longer ranged exchange interactions are antiferromagnetic \cite{Ferriani2008Jul, Bruning2022Jun, Rozsa2015Apr}. For computational convenience we replace a frustrated exchange interaction by a weak ferromagnetic exchange interaction limited to nearest neighbors. 
There exist theoretical proposals for how to tune $D$, $J$, and their ratio using lasers \cite{DMIguideSmallJ, ControlSmallJ, ExchangeFreeJ0}, potentially making helical states with arbitrary periodicities possible. 

The experiments in Refs.~\cite{Ferriani2008Jul, Bruning2022Jun} consider a magnetic monolayer of Mn grown on a W(001) substrate. This is a square lattice, but we do not believe our main results are limited to triangular lattices. Hence, by growing a conductor with negligible SOC on top of the Mn monolayer, the model we consider in this paper could be realized at the interface. This requires that the magnetic monolayer does not diffuse into the conductor forming alloys. Additionally, the conductor should not become superconducting due to electron-phonon interactions, such that any observed superconductivity is due to electron-magnon coupling. Alternatively an increase of $T_c$ in a low-$T_c$ phonon-driven superconductor could also confirm the presence of a magnon-driven pairing mechanism.

\subsection{Magnons} \label{sec:magnon}

To study spin fluctuations around the GS in the magnet using magnons, we perform a Holstein-Primakoff (HP) transformation \cite{HP-PhysRev.58.1098} adapted to noncolinear GSs \cite{HProtation_2009}. To this end, a site dependent orthonormal frame $\{\hat{e}_1^i, \hat{e}_2^i, \hat{e}_3^i\}$ with $\hat{e}_3^i = \boldsymbol{m}_i$ is introduced. Here, $\hat{e}_1^i = (\cos\theta_i\cos\phi_i, \cos\theta_i\sin\phi_i, -\sin\theta_i)$, $\hat{e}_2^i = (-\sin\phi_i, \cos\phi_i, 0)$, $\hat{e}_3^i = \boldsymbol{m}_i = (\sin\theta_i\cos\phi_i, \sin\theta_i\sin\phi_i, \cos\theta_i)$, with $\theta_i$ and $\phi_i$ the polar and azimuthal angle made by the spin at lattice site $i$ in the GS. The HP transformation is given by $S_{i3}=\boldsymbol{S}_i \cdot \hat{e}_3^i = S - a_i^\dagger a_i$, $S_{i\pm} = \boldsymbol{S}_i \cdot \hat{e}_\pm^i $, $\hat{e}_\pm^i = \hat{e}_1^i \pm i\hat{e}_2^i$, $S_{i+} \approx \sqrt{2S}a_i,$ and $S_{i-} \approx \sqrt{2S}a_i^\dagger$. Here, $a_i$ is an annihilation operator for a magnon at site $i$ related to spin flips along the local quantization axis. We have truncated at second order in magnons, assuming that the spin fluctuations are weak. This should be a good approximation at low temperatures \cite{HP-PhysRev.58.1098, HProtation_2009, QSkOP}.

The magnon operators are transformed to momentum space via the FT $a_i = \frac{1}{\sqrt{N'}} \sum_{\boldsymbol{q}\in \text{MBZ}} e^{i\boldsymbol{q}\cdot \boldsymbol{r}_i} a_{\boldsymbol{q}}^{(r)}$, assuming lattice site $i$ is located on sublattice $r$. Here, $N' = N/10$ is the number of sites on each sublattice. 

Inserting the HP and Fourier transformations into the Hamiltonian we get $H_{\text{MI}} = H_0 + H_1 + H_2$, where the subscripts indicate the order in magnon operators. $H_0$ is independent of magnon operators and corresponds to the classical Hamiltonian. $H_1$ contains terms that are linear in magnon operators. These terms vanish when expanding around the magnetic GS \cite{HProtation_2009}. The quadratic terms may be written as
\begin{equation}
\label{eq:H2}
    H_2 = \frac{1}{2}\sum_{\boldsymbol{q}}  \boldsymbol{a}_{\boldsymbol{q}}^\dagger M_{\boldsymbol{q}}  \boldsymbol{a}_{\boldsymbol{q}},
\end{equation}
where $\boldsymbol{a}_{\boldsymbol{q}}^\dagger = (a_{\boldsymbol{q}}^{(1)\dagger}, a_{\boldsymbol{q}}^{(2)\dagger}, \dots, a_{\boldsymbol{q}}^{(10)\dagger}, a_{-\boldsymbol{q}}^{(1)}, \dots, a_{-\boldsymbol{q}}^{(10)})$ and
\begin{equation}
\label{eq:M}
    M_{\boldsymbol{q}} = \begin{pmatrix} \eta_{\boldsymbol{q}} & \nu_{-\boldsymbol{q}}^* \\ \nu_{\boldsymbol{q}} & \eta_{-\boldsymbol{q}}^*    \end{pmatrix}.
\end{equation}
The matrix elements can be written as
\begin{align}
    \eta_{\boldsymbol{q}}^{r,s} =& \eta_r \delta_{r,s} +S\Lambda_{\boldsymbol{q}+}^{r,s},
\end{align}
\begin{align}
    \eta_r =& 2S\sum_s [J_{\boldsymbol{0}}^{(r,s)}\hat{e}_3^r \cdot \hat{e}_3^s -\boldsymbol{D}_{\boldsymbol{0}}^{(r,s)} \cdot (\hat{e}_3^r \cross \hat{e}_3^s)] \nonumber\\
    &-KS[1-3(\hat{e}_3^r \cdot \hat{z})^2] ,
\end{align}
\begin{align}
    \nu_{\boldsymbol{q}}^{r,s} =& \nu_r \delta_{r,s} +S\Lambda_{\boldsymbol{q}-}^{r,s} ,
\end{align}
\begin{align}
    \nu_r =& -KS(\hat{e}_1^r \cdot \hat{z})^2 ,
\end{align}
\begin{align}
    \Lambda_{\boldsymbol{q}\pm}^{r,s} =& -J_{\boldsymbol{q}}^{(r,s)}\hat{e}_{\pm}^r \cdot \hat{e}_-^s + \boldsymbol{D}_{\boldsymbol{q}}^{(r,s)} \cdot (\hat{e}_{\pm}^r \cross \hat{e}_-^s) .
\end{align}
Given that there exists $i\in r$ and $j\in s$ such that $i$ and $j$ are nearest neighbors, $J_{\boldsymbol{q}}^{(r,s)} = J\sum_{\boldsymbol{\delta}_{(r,s)}}e^{i\boldsymbol{q}\cdot \boldsymbol{\delta}_{(r,s)}}$ and $\boldsymbol{D}_{\boldsymbol{q}}^{(r,s)} = D\sum_{\boldsymbol{\delta}_{(r,s)}}e^{i\boldsymbol{q}\cdot \boldsymbol{\delta}_{(r,s)}} \boldsymbol{\delta}_{(r,s)} \cross \hat{z} $. If not, $J_{\boldsymbol{q}}^{(r,s)} = 0$ and $\boldsymbol{D}_{\boldsymbol{q}}^{(r,s)} = \boldsymbol{0}$. For 1D DMI, $\boldsymbol{D}_{\boldsymbol{q}}^{(r,s)} = \boldsymbol{0}$ also if $\boldsymbol{\delta}_{(r,s)} \neq \pm \hat{x}$. Here, $\boldsymbol{\delta}_{(r,s)}$ is the set of unit vectors from site $i\in r$ to sites $j\in s$ where $i$ and $j$ are nearest neighbors. With the sublattice numbering given in Fig.~\ref{fig:setupBZGS}(c) we have, e.g., $\boldsymbol{\delta}_{(0,6)} = \{ (1/2,\sqrt{3}/2),$ $(1/2, -\sqrt{3}/2)\}$.

This quadratic Hamiltonian is diagonalized using a matrix generalization of the Bogoliubov transformation \cite{COLPA}, giving
\begin{equation}
    H_2 = \sum_{\boldsymbol{q}}\sum_{n = 1}^{10} \omega_{\boldsymbol{q}n} \pqty{b_{\boldsymbol{q}n}^\dagger b_{\boldsymbol{q}n} + \frac{1}{2}},
\end{equation}
where $\omega_{\boldsymbol{q}n}$ is magnon band number $n$, the diagonalized magnon operators are given by $\boldsymbol{b}_{\boldsymbol{q}} = T_{\boldsymbol{q}}\boldsymbol{a}_{\boldsymbol{q}},$
and the paraunitary matrix
\begin{equation}
\label{eq:ColpaT}
    T_{\boldsymbol{q}} = \begin{pmatrix} U_{\boldsymbol{q}} & V_{-\boldsymbol{q}}^* \\ V_{\boldsymbol{q}} & U_{-\boldsymbol{q}}^*    \end{pmatrix}
\end{equation}
obeys $T_{\boldsymbol{q}}^{-1} = \mathcal{J}T_{\boldsymbol{q}}^\dagger \mathcal{J}$, with $\mathcal{J} = \operatorname{diag}(1,1,\dots, -1,-1, \dots)$.

\begin{figure}
    \centering
    \includegraphics[width=\linewidth]{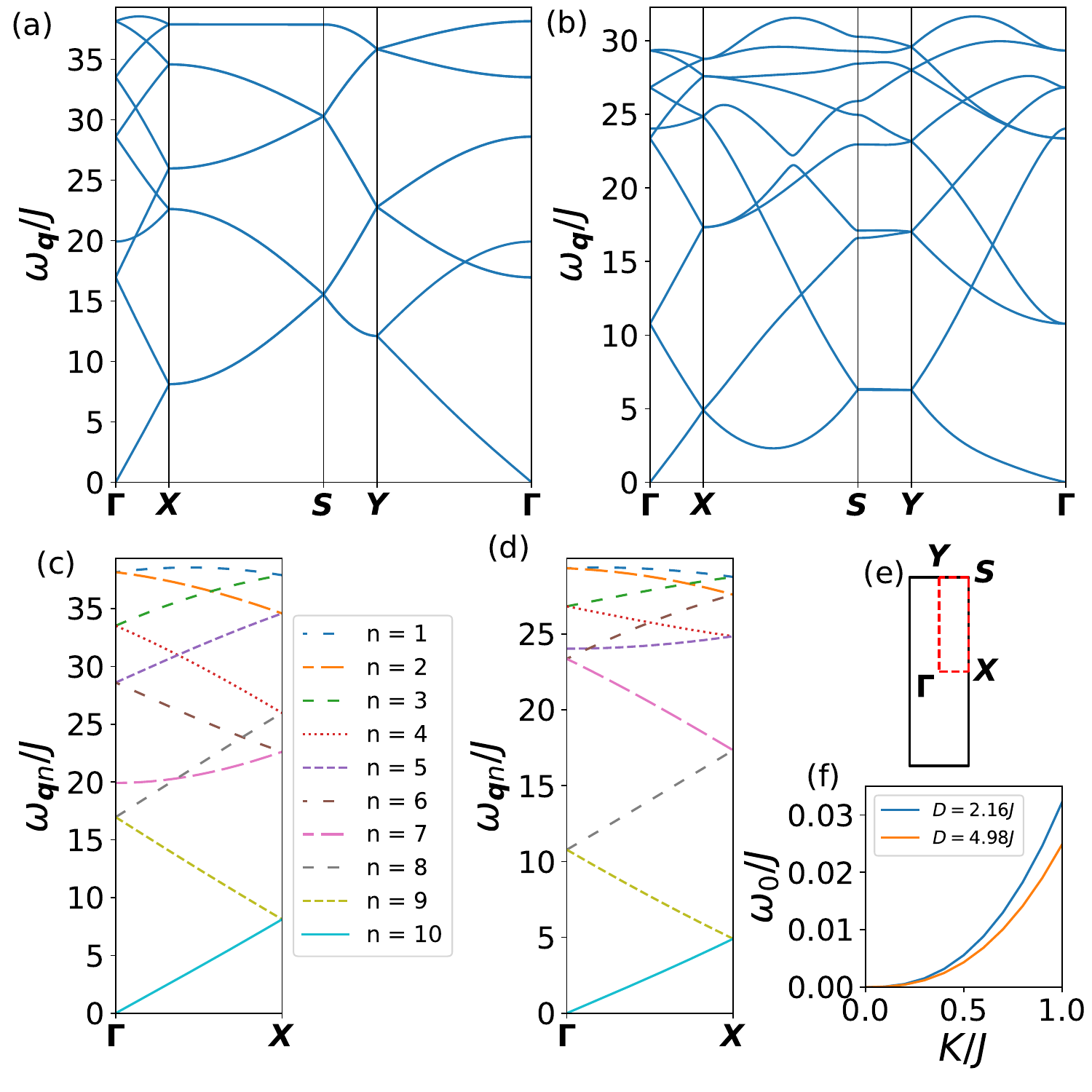}
    \caption{(a,b) Magnon spectrum for the GS shown in Fig.~\ref{fig:setupBZGS}(c) at $K = 0$, $S=1$, (a) 1D DMI with $D/J = 4.98$, and (b) full DMI with $D/J = 2.16$. The magnon bands are plotted along the path in the MBZ sketched in (e). Note that full DMI lifts several degeneracies in the magnon spectrum. (c,d) Magnon bands sorted by eigenvector overlap in the direction $\boldsymbol{\Gamma}-\boldsymbol{X}$ revealing several band crossings. (c) 1D DMI, (d) full DMI. (f) The gap in the magnon spectrum as a function of easy-axis anisotropy for both DMI models. The value of $D/J$ specifies the type of DMI. \label{fig:omq}}
\end{figure}

The ten magnon bands are shown in Fig.~\ref{fig:omq}. With 1D DMI we find paths in the MBZ where several bands are degenerate. By considering the overlap of eigenvectors \cite{BenestadMaster, AbnarMaster}, we are able to confirm several band crossings as well, see Figs.~\ref{fig:omq}(c) and (d). In either DMI model, no magnon bands are isolated, and so the magnons are topologically trivial \cite{ChernBoson1, ChernBoson2, ChernBosonTMI, TopoMagnonRev, QSkQTPT}. This is in contrast to the skyrmionic magnons studied in Refs.~\cite{QSkQTPT, Maeland2023PRL} where several magnon bands are topologically nontrivial. At zero easy-axis anisotropy and with all spins organized in the $xz$ plane, the MI Hamiltonian has a continuous symmetry. A rotation of all spins in the $xz$ plane by the same angle, preserves both the dot product and cross product of all spins. As a result there is a Goldstone mode, $\omega_0 = \min \omega_{\boldsymbol{q}n} = 0$. An easy-axis anisotropy breaks the continuous symmetry, introducing a magnon gap. Figure \ref{fig:omq}(f) shows the magnon gap for both DMI models as a function of $K$.

\subsection{Electron-magnon coupling}
The interfacial exchange interaction \cite{KargarianFMTI, HugdalTIFMAFM,  ArneFMNM,ArneAFMNM_Umklapp,EirikNMAFM, EirikTIAFM, EirikEliashberg, brekke2023interfacial, Brekke2023Aug, SkTopoSCNagaosa, SkTopoSCMajoranaChen, SkTopoSCMajoranaLossFM, SkMajoranaRex, SkTopoSCMajoranaLossAFM, SkTopoSCMajoranaDagotto, SkTopoSCgarnier, SkTopoSCMajoranaMascot, ExpMagnonInducedHeterostruct, ExpSkHeterostructure, ExpInterfaceExchange, Maeland2023PRL} is modeled by
\begin{equation}
    H_{\text{int}} = -2\Bar{J}\sum_i \boldsymbol{c}_i^\dagger \boldsymbol{\sigma} \boldsymbol{c}_i \cdot \boldsymbol{S}_i,
\end{equation}
where $\boldsymbol{c}_i = (c_{i\uparrow}, c_{i\downarrow})^T$ and $\boldsymbol{\sigma}$ is a vector containing the Pauli matrices. Applying the HP transformation yields \cite{Maeland2023PRL}
\begin{align}
\label{eq:Hem_i}
    H_{\text{int}} =& -2\Bar{J}\sqrt{\frac{S}{2}} \sum_{i\sigma} [e^{-i\sigma\phi_i}(\cos\theta_i-\sigma)a_i c_{i\sigma}^\dagger c_{i,-\sigma} + \text{H.c.}] \nonumber \\
    &+2\Bar{J}\sqrt{\frac{S}{2}}\sum_{i\sigma}(\sigma \sin\theta_i a_i c_{i\sigma}^\dagger c_{i\sigma} + \text{H.c.}) \nonumber \\
    &-2\Bar{J}S\sum_{i\sigma}[ \sin\theta_i e^{-i\sigma\phi_i} c_{i\sigma}^\dagger c_{i,-\sigma} +\sigma \cos\theta_i c_{i\sigma}^\dagger c_{i\sigma}].
\end{align}
Here, H.c.~denotes the Hermitian conjugate of the preceding term, $\sigma=1$ for spin up, and $\sigma = -1$ for spin down.
The terms in the last line are independent of magnon operators and would represent renormalizations of the electron spectrum of order $\Bar{J}/t$. We will consider $\Bar{J}/t \ll 1$ and neglect such renormalizations, as discussed in Appendix \ref{app:Higher}. Since the MBZ is smaller than the EBZ, Umklapp processes need to be included when applying a FT to Eq.~\eqref{eq:Hem_i} \cite{AbnarMaster, Maeland2023PRL}. We choose ten reciprocal lattice vectors $\boldsymbol{Q}_{\nu} = 2\pi\{(0,0),$ $ (\pm 1/5, 0),$ $ (\pm 2/5, 0),$$ (\pm 3/5, 0),$ $ (0, 1/\sqrt{3}),$ $ (\pm 1/5, 1/\sqrt{3})\}.$ By periodicity, the ten MBZs centered around these reciprocal lattice vectors cover the EBZ. The FT of the electron operators is rewritten as 
\begin{equation}
    c_{i\sigma} = \frac{1}{\sqrt{N}} \sum_{\boldsymbol{k}\in \text{MBZ}} \sum_\nu c_{\boldsymbol{k}+\boldsymbol{Q}_\nu, \sigma} e^{i(\boldsymbol{k}+\boldsymbol{Q}_\nu) \cdot\boldsymbol{r}_i} .
\end{equation}
Applying this FT and inserting diagonalized magnon operators yields \cite{Maeland2023PRL}
\begin{align}
    H_{\text{int}} =& \sum_{\substack{\boldsymbol{k}\in \text{EBZ} \\ \boldsymbol{q}\in \text{MBZ}}} \sum_{\nu, r,n,\sigma,\sigma'}  \Big[ g_{\nu r}^{\sigma\sigma'}\pqty{U_{\boldsymbol{q},r,n}^\dagger b_{\boldsymbol{q}n} -V_{\boldsymbol{q},r,n}^\dagger b_{-\boldsymbol{q}n}^\dagger } \nonumber \\ 
    &\times c_{\boldsymbol{k}+\boldsymbol{q}+\boldsymbol{Q}_\nu,\sigma}^\dagger c_{\boldsymbol{k},\sigma'} + \text{H.c.} \Big]
\end{align}
with
\begin{equation}
\label{eq:gunpol}
    g_{\nu r}^{\sigma,-\sigma} = -2\Bar{J}\sqrt{\frac{S}{2}} \frac{\sqrt{N'}}{N}e^{-i\sigma\phi_r}(\cos\theta_r-\sigma)  e^{-i\boldsymbol{Q}_{\nu} \cdot \boldsymbol{r}_r},
\end{equation}
\begin{equation}
\label{eq:gpol}
    g_{\nu r}^{\sigma,\sigma} = 2\Bar{J}\sqrt{\frac{S}{2}} \frac{\sqrt{N'}}{N}\sigma \sin\theta_r  e^{-i\boldsymbol{Q}_{\nu} \cdot \boldsymbol{r}_r}.
\end{equation}
For a helical GS in the $xz$ plane, $e^{-i\sigma\phi_r} = \cos(\phi_r)$ since $\phi_r = 0$ or $\pi$. The diagonalized magnons $b_{\boldsymbol{q}n}$ are linear combinations of the original sublattice magnons $a_{\boldsymbol{q}}^{(r)}$, which have their spin quantized along the spin direction of sublattice $r$. As a result, $b_{\boldsymbol{q}n}$ do not have well-defined quantization axes for spin, and therefore can be involved both in processes where the $z$-component of the electron spin flips, and processes where it remains unchanged. This is in contrast to colinear states \cite{ArneFMNM, ArneAFMNM_Umklapp, EirikNMAFM}, where a single magnon can only be involved in processes that flip electron spins. This distinction is a key component of noncolinear magnetic states, that later makes it possible to generate spin polarized Cooper pairs. 
In some systems with colinear magnetic states, electron-magnon scatterings involving a single magnon are negligible in the context of superconductivity. Under such circumstances, higher-order processes dominate and it is possible to generate spin polarized Cooper pairs also from colinear magnetic states \cite{Brekke2023Aug}.

\begin{figure*}
    \centering
    \includegraphics[width=0.9\linewidth]{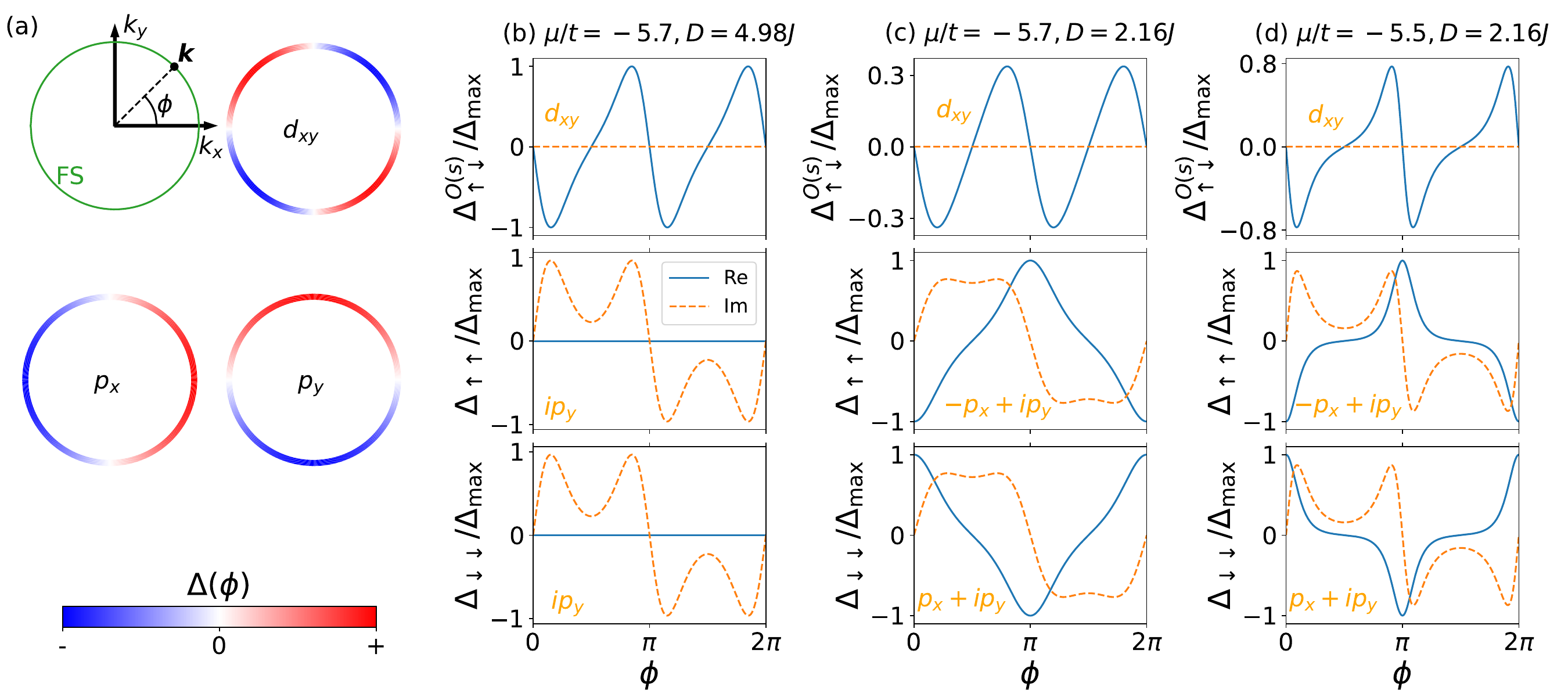}
    \caption{(a) A sketch of the FS in green with a definition of the angle $\phi$ that $\boldsymbol{k}$ makes with the $k_x$ axis. Red and blue indicate the sign of the relevant gap symmetries, assuming the gap is nonzero in a small region around the FS. (b) Solution to the linearized gap equation for 1D DMI, $D/J = 4.98$. $\Delta_{\boldsymbol{k}\uparrow\downarrow}^{E(s)} = 0$ is not shown. (c) Solution to the linearized gap equation for full DMI, $D/J = 2.16$, now showing a topologically nontrivial gap structure in the spin-polarized gaps. $\Delta_{\boldsymbol{k}\uparrow\downarrow}^{E(s)}/\Delta_{\text{max}} \approx 0$ is not shown. $\Delta_{\text{max}}$ is the amplitude of the largest real or imaginary part of any of the four gaps. (d) Same as (c) at higher chemical potential. The parameters are $t/J = 1000$, $K/J = 0.1$, $S  =1$, while $\Bar{J}$ does not affect the structure of the gap. \label{fig:Delta}}
\end{figure*}

\section{Superconductivity}
Applying a Schrieffer-Wolff transformation \cite{SchriefferWolff, Maeland2023PRL} to $H_{\text{int}}$, we get an effective electron-electron interaction $H_{\text{ee}}$,
\begin{widetext}
\begin{equation}
\label{eq:pairing}
    H_{\text{ee}} = \sum_{\substack{\boldsymbol{k}\boldsymbol{k}' \in \text{EBZ}\\ \boldsymbol{q} \in \text{MBZ}}} \sum_{ \substack{\nu\nu' \\\alpha\alpha'\beta\beta'}} V_{\boldsymbol{k}\boldsymbol{q}\nu\nu'}^{\alpha\alpha'\beta\beta'} c_{\boldsymbol{k}+\boldsymbol{q}+\boldsymbol{Q}_\nu,\alpha}^\dagger c_{\boldsymbol{k},\alpha'}c_{\boldsymbol{k}'-\boldsymbol{q}+\boldsymbol{Q}_{\nu'},\beta}^\dagger c_{\boldsymbol{k}',\beta'},
\end{equation}
with
\begin{align}
\label{eq:Vkq}
    V_{\boldsymbol{k}\boldsymbol{q}\nu\nu'}^{\alpha\alpha'\beta\beta'} = \sum_n \Big[ \frac{A_{\boldsymbol{q}n\nu\nu'}^{\alpha\alpha'\beta\beta'}}{\epsilon_{\boldsymbol{k}}-\epsilon_{\boldsymbol{k}+\boldsymbol{q}+\boldsymbol{Q}_\nu}+\omega_{\boldsymbol{q}n}} -\frac{B_{\boldsymbol{q}n\nu\nu'}^{\alpha\alpha'\beta\beta'}}{\epsilon_{\boldsymbol{k}}-\epsilon_{\boldsymbol{k}+\boldsymbol{q}+\boldsymbol{Q}_\nu}-\omega_{-\boldsymbol{q},n}} \Big],
\end{align}
\begin{align}
    A_{\boldsymbol{q}n\nu\nu'}^{\alpha\alpha'\beta\beta'} =& -\frac{1}{2} \sum_{rr'}\Big[ g_{\nu r}^{\alpha \alpha'} g_{\nu' r'}^{\beta \beta'} U_{\boldsymbol{q},r,n}^\dagger (-V_{-\boldsymbol{q},r',n}^\dagger)+g_{\nu r}^{\alpha \alpha'} g_{\bar{\nu}' r'}^{\beta' \beta*} U_{\boldsymbol{q},r,n}^\dagger U_{\boldsymbol{q},r',n}^T  \nonumber \\
    &+g_{\bar{\nu} r}^{\alpha' \alpha*} g_{\nu' r'}^{\beta \beta'} (-V_{-\boldsymbol{q},r,n}^T) (-V_{-\boldsymbol{q},r',n}^\dagger)  +g_{\bar{\nu} r}^{\alpha' \alpha*} g_{\bar{\nu}' r'}^{\beta' \beta*} (-V_{-\boldsymbol{q},r,n}^T) U_{\boldsymbol{q},r',n}^T \Big],
\end{align}
\end{widetext}
and $B_{\boldsymbol{q}n\nu\nu'}^{\alpha\alpha'\beta\beta'} =  A_{-\boldsymbol{q}n\nu'\nu}^{\beta\beta'\alpha\alpha'}$. $\Bar{\nu}$ is defined such that $\boldsymbol{Q}_{\Bar{\nu}} = -\boldsymbol{Q}_\nu$ modulo a reciprocal lattice vector for the triangular lattice. The effective electron-electron interaction is mediated by the magnons of the MI and it depends both on the magnetic GS and the transformation matrix used to diagonalize the magnons.

We assume zero net momentum Cooper pairs and symmetrize the interaction \cite{Sigrist, Maeland2023PRL}
\begin{equation}
\label{eq:Vbargen}
    H_{\text{ee}} = \frac{1}{2} \sum_{\boldsymbol{k}\boldsymbol{k}'\in \text{EBZ}} \sum_{\{\sigma_i\}} \bar{V}_{\boldsymbol{k}\boldsymbol{k}'}^{\sigma_1 \sigma_2 \sigma_3 \sigma_4} c_{\boldsymbol{k}',\sigma_1}^\dagger c_{-\boldsymbol{k}',\sigma_2}^\dagger c_{-\boldsymbol{k},\sigma_3}c_{\boldsymbol{k},\sigma_4}.
\end{equation}
Here, $\bar{V}_{\boldsymbol{k}\boldsymbol{k}'}^{\sigma_1 \sigma_2 \sigma_3 \sigma_4} = (V_{\boldsymbol{k}\boldsymbol{k}'}^{\sigma_1 \sigma_2 \sigma_3 \sigma_4}+V_{-\boldsymbol{k},-\boldsymbol{k}'}^{\sigma_2 \sigma_1 \sigma_4 \sigma_3}-V_{-\boldsymbol{k},\boldsymbol{k}'}^{\sigma_1 \sigma_2 \sigma_4 \sigma_3}-V_{\boldsymbol{k},-\boldsymbol{k}'}^{\sigma_2 \sigma_1 \sigma_3 \sigma_4})/2$ and $V_{\boldsymbol{k}\boldsymbol{k}'}^{\sigma_1 \sigma_2 \sigma_3 \sigma_4} = V_{\boldsymbol{k}\boldsymbol{q}\nu\Bar{\nu}}^{\sigma_1 \sigma_4 \sigma_2 \sigma_3}$. The momentum $\boldsymbol{q} \in \text{MBZ}$ and Umklapp process $\nu$ are found from $\boldsymbol{k}'-\boldsymbol{k}$. First, $\nu$ is determined from which $\boldsymbol{Q}_\nu$ is closest to $\boldsymbol{k}'-\boldsymbol{k}$ modulo a reciprocal lattice vector for the triangular lattice. Then, $\boldsymbol{q} = \boldsymbol{k}'-\boldsymbol{k}-\boldsymbol{Q}_\nu$.

We now treat the electron-electron interaction using mean-field theory, and define a superconducting gap function $\Delta_{\boldsymbol{k}\sigma_1\sigma_2} = -\sum_{\boldsymbol{k}'\sigma_3\sigma_4} \bar{V}_{\boldsymbol{k}'\boldsymbol{k}}^{\sigma_1 \sigma_2 \sigma_3 \sigma_4} \langle c_{-\boldsymbol{k}',\sigma_3}c_{\boldsymbol{k}',\sigma_4} \rangle $ \cite{BenestadMaster, AbnarMaster, Maeland2023PRL, Sigrist}. Since all $\bar{V}_{\boldsymbol{k}'\boldsymbol{k}}^{\sigma_1 \sigma_2 \sigma_3 \sigma_4}$ are nonzero in general, all spin combinations in the gaps are possible and they can coexist. The spin singlet gap is defined as $\Delta_{\boldsymbol{k}\uparrow\downarrow}^{O(s)} = (\Delta_{\boldsymbol{k}\uparrow\downarrow}-\Delta_{\boldsymbol{k}\downarrow\uparrow})/2$ and the spin triplet gaps are $\Delta_{\boldsymbol{k}\uparrow\uparrow}$, $\Delta_{\boldsymbol{k}\uparrow\downarrow}^{E(s)} = (\Delta_{\boldsymbol{k}\uparrow\downarrow}+\Delta_{\boldsymbol{k}\downarrow\uparrow})/2$, and  $\Delta_{\boldsymbol{k}\downarrow\downarrow}$. 
With details relegated to Appendix \ref{app:gapeq}, we obtain a linearized and Fermi surface (FS) averaged gap equation
\begin{equation}
\label{eq:linFSavrgap}
   \lambda \boldsymbol{\Delta}(\phi) = -N_0 \langle \mathcal{V}(\phi', \phi) \boldsymbol{\Delta}(\phi') \rangle_{\text{FS},\phi'},
\end{equation}
where $N_0 = \sum_{\boldsymbol{k}}\delta(\epsilon_{\boldsymbol{k}})$ is the density of states per spin on the FS and $\phi$ is the angle $\boldsymbol{k}$ makes with the $k_x$ axis. The gaps are organized in a vector $\boldsymbol{\Delta}(\phi) = [\Delta_{\uparrow\downarrow}^{O(s)}(\phi), \Delta_{\uparrow\uparrow}(\phi), \Delta_{\downarrow\downarrow}(\phi), \Delta_{\uparrow\downarrow}^{E(s)}(\phi)]^T$ and $\mathcal{V}(\phi', \phi)$ is a matrix of coupling elements defined in Eq.~\eqref{eq:V44}. Equation \eqref{eq:linFSavrgap} is solved as an eigenvalue problem. The largest eigenvalue $\lambda$ provides an estimate of the critical temperature
\begin{equation}
    k_B T_c \approx 1.13\omega_c e^{-1/\lambda},
\end{equation}
where $\omega_c = \max \omega_{\boldsymbol{q}n}$ is the maximum magnon energy.
The eigenvector corresponding to $\lambda$ gives information about the angular dependence of the gap function on the FS. Due to the exponential dependence on $\lambda$, $T_c$ is extremely sensitive to the chosen material parameters. Therefore, this BCS estimate of the critical temperature is considered nonpredictive, serving merely as a rough guide of what to expect for $T_c$. On the other hand, BCS theory does predict an accurate value of the ratio of the gap amplitude at zero temperature $\Delta_0$ and the critical temperature $T_c$, namely $2\Delta_0/k_B T_c \approx 3.53$, which holds well for many weak-coupling conventional superconductors \cite{SFsuperconductivity}.

\subsection{Solutions of the linearized gap equation}
Solutions of the linearized gap equation are shown in Fig.~\ref{fig:Delta}. With 1D DMI we get that $\Delta_{\boldsymbol{k}\uparrow\downarrow}^{O(s)}$ can be chosen to be real and displays $d_{xy}$-wave symmetry. Then, the spin polarized gaps are purely imaginary, with $ip_y$-wave symmetry. The unpolarized spin triplet gap is zero, $\Delta_{\boldsymbol{k}\uparrow\downarrow}^{E(s)} = 0$. There is no bulk gap in the spectrum $E_{\boldsymbol{k}\eta}$ [Eq.~\eqref{eq:SCbands}] in the superconducting state, and we therefore conclude that there is no strong TSC in the case of 1D DMI.

With full DMI, $\Delta_{\boldsymbol{k}\uparrow\downarrow}^{O(s)}$ shows $d_{xy}$-wave symmetry, while $\Delta_{\boldsymbol{k}\downarrow\downarrow}$ shows $p_x+ip_y$-wave symmetry, known as chiral $p$ wave, where there is a phase difference between the real and imaginary part. $\Delta_{\boldsymbol{k}\uparrow\uparrow} = -\Delta_{\boldsymbol{k}\downarrow\downarrow}^*$ is also chiral $p$ wave. $\Delta_{\boldsymbol{k}\uparrow\downarrow}^{E(s)}$ has a negligible amplitude compared to the other gaps. Such a gap solution can be shown to be time-reversal symmetric (TRS) \cite{Sigrist, Maeland2023PRL}. 
There is a bulk gap in the spectrum $E_{\boldsymbol{k}\eta}$ and in the next section we compute the topological invariant and find that the system is a topological superconductor. Note that the above gap symmetries are the same gap symmetries obtained in the chiral $p$-wave phase of Ref.~\cite{Maeland2023PRL}. For full DMI, the gap symmetries are the same for all parameters in the region $\mu/t \in (-6.0, -5.43]$ and $K/J \in (0,1]$.

\begin{figure}
    \centering
    \includegraphics[width=\linewidth]{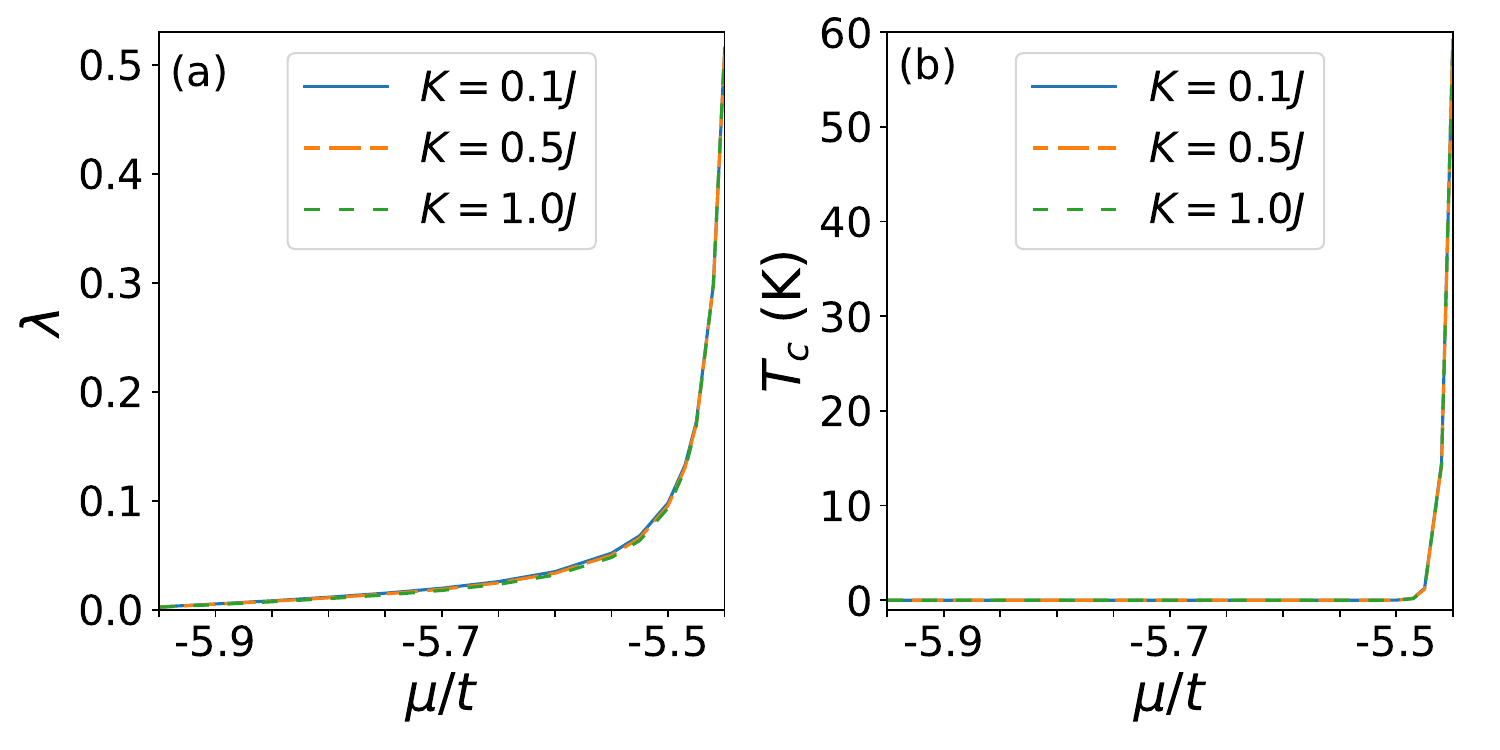}
    \caption{(a) $\lambda$ and (b) $T_c$ as a function of chemical potential for some chosen easy-axis anisotropy strengths using full DMI. Note the enhancement as we approach the chemical potential where the FS reaches the edge of the MBZ, i.e. where the Fermi momentum $k_F \approx \pi/5$. Also note the weak dependence on $K/J$. The parameters are $t/J = 1000$, $\Bar{J}/J = 50$, $D/J = 2.16$, and $S=1$. To find $T_c$ in units of K, $J = 1$~meV was chosen.}
    \label{fig:lam}
\end{figure}

When the FS approaches the edges of the MBZ, i.e., when the Fermi momentum in the $k_x$ direction $k_F \approx \pi/5$ at $\mu/t \approx -5.43$, we find an enhancement of superconductivity. Figure \ref{fig:lam} shows $\lambda$ and $T_c$ as functions of the chemical potential. We attribute this increase of $T_c$ to Umklapp enhancement. This is analogous to the case of antiferromagnet/NM heterostructures. The reader is referred to Refs.~\cite{Sun2023Aug, EirikEliashberg} for analytic expressions showing the mechanism for Umklapp enhancement. Here, we do not have analytic expressions for the magnon transformation matrix elements, due to the size of the matrices. Due to the Bose statistics of magnons, these matrix elements are not bounded from above. Without Umklapp, large matrix elements tend to cancel. For very small FSs where there are no Umklapp processes, we find that $A_{\boldsymbol{q}n\nu\nu'}^{\alpha\alpha'\beta\beta'}$ in Eq.~\eqref{eq:Vkq} is null close to zero momentum for the lowest energy magnon. For larger FSs with Umklapp effects, $A_{\boldsymbol{q}n\nu\nu'}^{\alpha\alpha'\beta\beta'}$ can become large in magnitude at zero momentum also for the lowest energy magnon. This results in a stronger interaction, leading to an enhancement of superconductivity. In the limit $k_F \to \pi/5$ the interaction becomes so strong that weak-coupling theory of superconductivity presumably becomes unreliable, and a strong-coupling approach using Eliashberg theory should be preferred \cite{EirikEliashberg}. A similar Umklapp enhancement was not found in the case of a skyrmion GS in the MI in Ref.~\cite{Maeland2023PRL}, presumably due to the fact that the gap transitions to chiral $f$-wave symmetry before the FS approaches the edges of the MBZ. Hence, helical states may be better candidates than skyrmions when searching for magnon-induced TSC in MI/NM heterostructures since they could generate higher $T_c$.

\begin{figure}
    \centering
    \includegraphics[width=\linewidth]{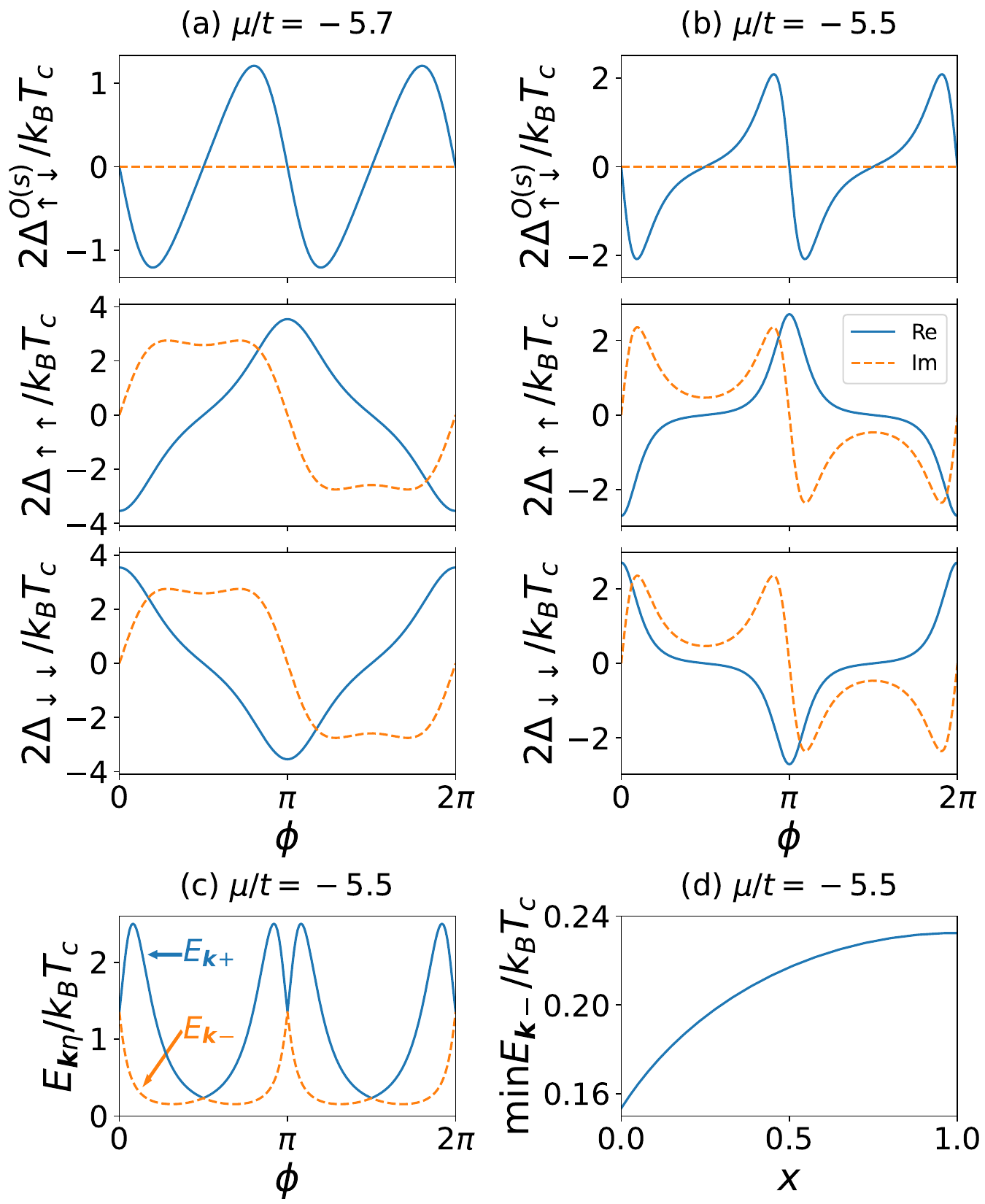}
    \caption{(a,b) The zero temperature gap compared to $k_B T_c$ for full DMI. A coexisting $\Delta_{\boldsymbol{k}\uparrow\downarrow}^{E(s)}$ gap with negligible amplitude is not shown. The gap structure is strikingly similar to that at $T_c$ shown in Figs.~\ref{fig:Delta}(c) and (d). In (a), for $\mu/t = -5.7$, the largest gap amplitude compares favorably to the BCS prediction $2\Delta_0/k_B T_c \approx 3.53$. For a larger FS in (b), where Umklapp effects become important, $T_c$ is more enhanced than the gap amplitude at zero temperature. (c) shows the energy spectrum in the superconducting state, defined in Eq.~\eqref{eq:SCbands}. In (d), the minimum of the lowest band $E_{\boldsymbol{k}-}$ is plotted when the zero net spin gaps $\Delta_{\boldsymbol{k}\uparrow\downarrow}^{O(s)}$ and $\Delta_{\boldsymbol{k}\uparrow\downarrow}^{E(s)}$ are removed in a continuous fashion as a function of $x$. The parameters are $t/J = 1000$, $D/J = 2.16$, $K=0.1$ and $S=1$. The results are weakly dependent on $\bar{J}$. }
    \label{fig:DT0}
\end{figure}

\section{Topological superconductivity}

Strong TSC requires a bulk gap. Hence, the gap amplitude must be nonzero, and we must move below $T_c$. References \cite{Maeland2023PRL, Sun2023Aug} indicate that gap solutions at $T_c$ generally remain stable all the way down to zero temperature. A detailed derivation of the FS averaged zero temperature gap equation is given in the Supplemental Material of Ref.~\cite{Maeland2023PRL}. Figure \ref{fig:DT0} shows solutions to the zero temperature gap equation for full DMI. We find the same gap symmetry as at $T_c$, with an amplitude comparable to the BCS prediction $2\Delta_0/k_B T_c \approx 3.53$ in the case of $\mu/t = -5.7$. When approaching $\mu/t = -5.43$ Umklapp enhancement increases $T_c$ more than the zero temperature gap amplitude so that the gap amplitude is below the BCS estimate.

We intend to apply a topological invariant for TRS TSC that requires the spin up and spin down sector to decouple, as done in Ref.~\cite{Maeland2023PRL}. Hence, we must show that removing $\Delta_{\boldsymbol{k}\uparrow\downarrow}^{O(s)}$ and $\Delta_{\boldsymbol{k}\uparrow\downarrow}^{E(s)}$ in a continuous fashion does not close the bulk gap. We multiply $\Delta_{\boldsymbol{k}\uparrow\downarrow}$ and $\Delta_{\boldsymbol{k}\downarrow\uparrow}$ in the Bogoliubov-de Gennes (BdG) Hamiltonian in Eq.~\eqref{eq:BdG} by $1-x$. Tuning $x$ from 0 to 1 we do not observe any gap closings in the spectrum $E_{\boldsymbol{k}\eta}$, see Fig.~\ref{fig:DT0}(d). Hence, the present SC state is topologically equivalent to one where $\Delta_{\boldsymbol{k}\uparrow\downarrow}^{O(s)} = \Delta_{\boldsymbol{k}\uparrow\downarrow}^{E(s)} = 0$. We can then define two independent BdG Hamiltonians for each spin, $H_{\boldsymbol{k}\sigma} = \boldsymbol{d}_{\boldsymbol{k}\sigma} \cdot \boldsymbol{\sigma}$, with $\boldsymbol{d}_{\boldsymbol{k}\sigma} = (\Re\Delta_{\boldsymbol{k}\sigma\sigma}, -\Im\Delta_{\boldsymbol{k}\sigma\sigma}, \epsilon_{\boldsymbol{k}})$. The bulk topological invariant is defined as
\begin{equation}
\label{eq:topoinv}
    \nu_{\mathbb{Z}_2} = \frac{1}{2} (N_\uparrow - N_\downarrow)  \text{~mod~} 2,
\end{equation}
where $N_\sigma = (1/8\pi) \int_{\text{EBZ}} d \boldsymbol{k} \epsilon_{ij} \hat{d}_{\boldsymbol{k}\sigma} \cdot (\partial_{k_i} \hat{d}_{\boldsymbol{k}\sigma} \cross \partial_{k_j} \hat{d}_{\boldsymbol{k}\sigma})$ is a winding number, counting the number of times $\hat{d}_{\boldsymbol{k}\sigma}$ wraps the unit sphere. $\epsilon_{ij}$ is the Levi-Civita tensor, the indices $i$ and $j$ are summed over the set $\{x,y\}$, and $\hat{d}_{\boldsymbol{k}\sigma} = \boldsymbol{d}_{\boldsymbol{k}\sigma}/|\boldsymbol{d}_{\boldsymbol{k}\sigma}|$. $\nu_{\mathbb{Z}_2} = 0$ for a topologically trivial superconductor and $\nu_{\mathbb{Z}_2} = 1$ for a topologically nontrivial superconductor \cite{Bernevig2013}. Away from the FS, $\epsilon_{\boldsymbol{k}} \gg |\Delta_{\boldsymbol{k}\sigma\sigma}|$ and so the integrand of $N_\sigma$ is completely dominated by the behavior on the FS. Solutions of the FS averaged linearized gap equation is sufficient to compute $N_\sigma$. Using adaptive integration \cite{AdaptQuad} we obtain $N_\downarrow \approx 1, N_\uparrow \approx -1$ giving $\nu_{\mathbb{Z}_2} = 1$ when considering full DMI. Alternate definitions of the topological invariant based on FS topology \cite{Sato2009Jun, Sato2010Jun, Fu2010Aug, Qi2010Apr} would yield the same conclusion.

Having defined the topological invariant in Eq.~\eqref{eq:topoinv}, we now apply symmetry arguments to explain why 1D DMI does not yield strong TSC. 
For 1D DMI, the Hamiltonian and the GS have a $C_{2y}^s \mathcal{T}$ symmetry, where the operation $C_{2y}^s$ acts only in spin space and involves a rotation by $\pi$ around the $y$ axis, while $\mathcal{T}$ denotes time-reversal.
We then expect that $N_\uparrow$ is invariant when applying $C_{2y}^s \mathcal{T}$. On the other hand, applying $C_{2y}^s \mathcal{T}$ to the vector $\boldsymbol{d}_{\boldsymbol{k}\uparrow} = (\Re\Delta_{\boldsymbol{k}\uparrow\uparrow}, -\Im\Delta_{\boldsymbol{k}\uparrow\uparrow}, \epsilon_{\boldsymbol{k}})$ yields $(-\Re\Delta_{\boldsymbol{k}\uparrow\uparrow}, -\Im\Delta_{\boldsymbol{k}\uparrow\uparrow}, \epsilon_{\boldsymbol{k}})$, i.e., opposite winding. Hence, $N_\uparrow = -N_\uparrow$ which is only valid if $N_\uparrow = 0$. By the same argument $N_\downarrow = 0$, hence, $\nu_{\mathbb{Z}_2} = 0$. Thus, the superconducting state has trivial topology. Full DMI, on the other hand, breaks the $C_{2y}^s \mathcal{T}$ symmetry. 
The broken symmetry opens up the possibility for TSC.  
The presence of TSC must still be proved by solving the gap equations for specific systems.

In Appendix \ref{app:origin}, we discuss in more detail why the full DMI yields TSC in the specific system we study. We show that full DMI results in a sufficiently complicated electron-electron interaction to generate TSC. This shows that the TSC derived in Ref.~\cite{Maeland2023PRL} does not rely on topological magnons nor noncoplanar spins in the GS. Results in this paper show that topologically trivial magnons from coplanar magnetic states can also induce TSC in a normal metal through interfacial exchange interaction. This generalizes the results in Ref.~\cite{Maeland2023PRL} to a larger class of magnetic states. Noncolinear spins in the GS facilitate spin polarized Cooper pairs, while full DMI breaks enough symmetries to permit topologically nontrivial gap functions.

References \cite{Bernevig2013, Leijnse2012TSCrev, Ivanov2001Jan, Roy2008Mar} discuss the possible application of TRS TSC in topological quantum computation. This requires the generation of half-quantum vortices, which break TRS and affect only one spin component \cite{Leijnse2012TSCrev}. Such half-quantum vortices are possible in certain superconductors \cite{Ivanov2001Jan, Volovik1999Dec}. In general, superconducting vortices can be generated by magnetic fields or defects. MBSs would appear in the core of vortices due to the topologically nontrivial nature of the superconductor. MBSs can also appear if the gap is closed by variation in the chemical potential or by applying an electrostatic potential \cite{Leijnse2012TSCrev}. Reference \cite{Qi2009TRSTSC} discusses TRS vortices in TRS TSC. Such vortices would yield two MBSs in the core and would not be applicable in topological quantum computation \cite{Bernevig2013}. On the other hand, fusion of two TRS vortices could be used to demonstrate a condensed matter analog of supersymmetry \cite{Qi2009TRSTSC}.

Another key component of TSC for the application in topological quantum computation is the size of the bulk gap. In our proposed system, we found an enhancement of $T_c$ for certain chemical potentials. This results in a large gap amplitude at near zero temperature. However, the unconventional nature of the superconducting state yielded two bands $E_{\boldsymbol{k}\eta}$, defined in Eq.~\eqref{eq:SCbands} and plotted in Fig.~\ref{fig:DT0}(c). The topological gap is the minimum of $E_{\boldsymbol{k},-}$, which tends to be smaller than the gap amplitude, especially at large chemical potential, see Figs.~\ref{fig:DT0}(c) and (d). The Umklapp enhancement is strongest at $\phi = 0$ and $\pi$ [where the FS approaches the MBZ, see Fig.~\ref{fig:setupBZGS}(b)], and so the gap magnitudes are largest there. The gap magnitudes are relatively small at $\phi = \pi/2$ and $3\pi/2$. Nevertheless, the topological gap is enhanced due to the dramatic increase in $T_c$.

\section{Conclusion} \label{sec:Con}
We have studied the interface of a normal metal and a magnetic insulator with a coplanar, noncolinear helical ground state. Two models of the Dzyaloshinskii-Moriya interaction were applied and compared in terms of symmetry. We considered superconductivity induced in the normal metal due to an effective electron-electron interaction mediated by spin fluctuations in the magnetic state. With DMI active on all nearest-neighbor links, topological superconductivity is induced in the normal metal. Compared to previous study involving skyrmions, we find that noncoplanar spins in the magnetic ground state are not essential for the magnons to induce topological superconductivity. In a certain range of chemical potentials, Umklapp scattering causes an enhancement of the critical temperature of superconductivity in the case of helical magnetic states.

\section*{Acknowledgments}
We thank Bj{\o}rnulf Brekke for helpful discussions. We acknowledge funding from the Research Council of Norway (RCN) through its Centres of Excellence funding scheme Project No.~262633, ``QuSpin", and RCN through Project No.~323766, ``Equilibrium and out-of-equilibrium quantum phenomena in superconducting hybrids with antiferromagnets and topological insulators".

\appendix
\begin{widetext}
\section{Deriving the gap equation} \label{app:gapeq}
Cooper pair ensemble averages are defined as $b_{\boldsymbol{k}\sigma \sigma'} = \langle c_{-\boldsymbol{k},\sigma}c_{\boldsymbol{k}\sigma'}\rangle$. Inserting $c_{-\boldsymbol{k},\sigma}c_{\boldsymbol{k}\sigma'} = b_{\boldsymbol{k}\sigma \sigma'} + \delta b_{\boldsymbol{k}\sigma \sigma'}$, where the fluctuations around the mean-field value, $\delta b_{\boldsymbol{k}\sigma \sigma'}$, are assumed to be small and truncated to first order, the Hamiltonian can be written as
\begin{equation}
    H = \sum_{\boldsymbol{k}} \epsilon_{\boldsymbol{k}} + \frac{1}{2} \sum_{\boldsymbol{k}}\sum_{\sigma_1\sigma_2} \Delta_{\boldsymbol{k}\sigma_1 \sigma_2} \langle c_{\boldsymbol{k}\sigma_1}^\dagger c_{-\boldsymbol{k}\sigma_2}^\dagger \rangle + \frac{1}{2} \sum_{\boldsymbol{k}} \Psi_{\boldsymbol{k}}^\dagger H_{\boldsymbol{k}} \Psi_{\boldsymbol{k}},
\end{equation}
with $ \Psi_{\boldsymbol{k}} = (c_{\boldsymbol{k}\uparrow}, c_{\boldsymbol{k}\downarrow}, c_{-\boldsymbol{k}\downarrow}^\dagger, c_{-\boldsymbol{k}\uparrow}^\dagger)^T$, and the BdG Hamiltonian
\begin{equation}
\label{eq:BdG}
    H_{\boldsymbol{k}} = 
    \begin{pmatrix}
    \epsilon_{\boldsymbol{k}} & 0 & \Delta_{\boldsymbol{k}\uparrow\downarrow} & \Delta_{\boldsymbol{k}\uparrow\uparrow} \\
    0 & \epsilon_{\boldsymbol{k}} & \Delta_{\boldsymbol{k}\downarrow\downarrow} & \Delta_{\boldsymbol{k}\downarrow\uparrow} \\
    \Delta_{\boldsymbol{k}\uparrow\downarrow}^\dagger & \Delta_{\boldsymbol{k}\downarrow\downarrow}^\dagger & -\epsilon_{\boldsymbol{k}} & 0 \\
    \Delta_{\boldsymbol{k}\uparrow\uparrow}^\dagger & \Delta_{\boldsymbol{k}\downarrow\uparrow}^\dagger & 0 & -\epsilon_{\boldsymbol{k}}  \\
    \end{pmatrix}.
\end{equation}
The mean-field Hamiltonian is diagonalized, yielding 
\begin{align}
    H =& \sum_{\boldsymbol{k}} \epsilon_{\boldsymbol{k}}-\frac{1}{2}\sum_{\boldsymbol{k}\eta} E_{\boldsymbol{k}\eta} + \frac{1}{2} \sum_{\boldsymbol{k}}\sum_{\sigma_1\sigma_2} \Delta_{\boldsymbol{k}\sigma_1 \sigma_2}\langle c_{\boldsymbol{k}\sigma_1}^\dagger c_{-\boldsymbol{k}\sigma_2}^\dagger \rangle + \sum_{\boldsymbol{k}\eta} E_{\boldsymbol{k}\eta}\gamma_{\boldsymbol{k}\eta}^\dagger\gamma_{\boldsymbol{k}\eta}.
\end{align}
with the two bands in the superconducting state $E_{\boldsymbol{k}\pm}$ being
\begin{equation}
\label{eq:SCbands}
    E_{\boldsymbol{k}\pm} = \sqrt{\epsilon_{\boldsymbol{k}}^2 + \frac{1}{2}\Tr\hat{\Delta}_{\boldsymbol{k}} \hat{\Delta}_{\boldsymbol{k}}^\dagger \pm \frac{1}{2}\sqrt{A_{\boldsymbol{k}}}},
\end{equation}
with $[\hat{\Delta}_{\boldsymbol{k}}]_{\sigma_1\sigma_2} = {\Delta}_{\boldsymbol{k}\sigma_1\sigma_2}$,
\begin{equation}
    \frac{1}{2}\Tr\hat{\Delta}_{\boldsymbol{k}} \hat{\Delta}_{\boldsymbol{k}}^\dagger =\frac{1}{2} (|\Delta_{\boldsymbol{k}\uparrow\uparrow}|^2 + |\Delta_{\boldsymbol{k}\uparrow\downarrow}|^2 + |\Delta_{\boldsymbol{k}\downarrow\uparrow}|^2 + |\Delta_{\boldsymbol{k}\downarrow\downarrow}|^2   ),
\end{equation}
and
\begin{align}
    A_{\boldsymbol{k}} =& (|\Delta_{\boldsymbol{k}\uparrow\uparrow}|^2-|\Delta_{\boldsymbol{k}\downarrow\downarrow}|^2)^2 + (|\Delta_{\boldsymbol{k}\uparrow\downarrow}|^2 - |\Delta_{\boldsymbol{k}\downarrow\uparrow}|^2)^2 +2(|\Delta_{\boldsymbol{k}\uparrow\uparrow}|^2 +|\Delta_{\boldsymbol{k}\downarrow\downarrow}|^2)(|\Delta_{\boldsymbol{k}\uparrow\downarrow}|^2 + |\Delta_{\boldsymbol{k}\downarrow\uparrow}|^2)\nonumber \\ &+4\Delta_{\boldsymbol{k}\uparrow\uparrow}\Delta_{\boldsymbol{k}\downarrow\downarrow}\Delta_{\boldsymbol{k}\uparrow\downarrow}^\dagger \Delta_{\boldsymbol{k}\downarrow\uparrow}^\dagger + 4\Delta_{\boldsymbol{k}\uparrow\uparrow}^\dagger\Delta_{\boldsymbol{k}\downarrow\downarrow}^\dagger\Delta_{\boldsymbol{k}\uparrow\downarrow} \Delta_{\boldsymbol{k}\downarrow\uparrow}.
\end{align}
With the gap symmetries we obtain $ A_{\boldsymbol{k}}/16 = (\Delta_{\boldsymbol{k}\uparrow\downarrow}^{E(s)})^2 (\Delta_{\boldsymbol{k}\uparrow\downarrow}^{O(s)})^2 -\Delta_{\boldsymbol{k}\uparrow\uparrow}\Delta_{\boldsymbol{k}\downarrow\downarrow}(\Delta_{\boldsymbol{k}\uparrow\downarrow}^{O(s)})^2 $ making it clear that $A_{\boldsymbol{k}}$ is only nonzero for coexistence of spin singlet and spin triplet Cooper pairs.

Minimizing the free energy yields the gap equation \cite{AbnarMaster, Maeland2023PRL,Sigrist}. Introducing $B_{\boldsymbol{k}\sigma_1\sigma_2}^\dagger \equiv \frac{1}{4\sqrt{A_{\boldsymbol{k}}}}\pdv{A_{\boldsymbol{k}}}{\Delta_{\boldsymbol{k}\sigma_1\sigma_2}}$, $\boldsymbol{\Delta}_{\boldsymbol{k}} = (\Delta_{\boldsymbol{k}\uparrow\downarrow}^{O(s)}, \Delta_{\boldsymbol{k}\uparrow\uparrow}, \Delta_{\boldsymbol{k}\downarrow\downarrow}, \Delta_{\boldsymbol{k}\uparrow\downarrow}^{E(s)})^T$, and $\boldsymbol{B}_{\boldsymbol{k}} = (B_{\boldsymbol{k}\uparrow\downarrow}^{O(s)}, B_{\boldsymbol{k}\uparrow\uparrow}, B_{\boldsymbol{k}\downarrow\downarrow}, B_{\boldsymbol{k}\uparrow\downarrow}^{E(s)})^T$
we can write the gap equation as
\begin{equation}
    \label{eq:gapeqVector}
    \boldsymbol{\Delta_k} = -\sum_{\boldsymbol{k}'} \mathcal{V}_{\boldsymbol{k}'\boldsymbol{k}} \sum_\eta \pqty{\frac{1}{2}\boldsymbol{\Delta}_{\boldsymbol{k}'}+\eta \boldsymbol{B}_{\boldsymbol{k}'} }\chi_{\boldsymbol{k}'\eta},
\end{equation}
with $\chi_{\boldsymbol{k}\eta} = \tanh(\beta E_{\boldsymbol{k}\eta}/2)/2E_{\boldsymbol{k}\eta}$, 
\begin{equation}
\label{eq:V44}
    \mathcal{V}_{\boldsymbol{k}'\boldsymbol{k}} = \begin{pmatrix}
    \tilde{V}_{\boldsymbol{k}'\boldsymbol{k}}^{E(\boldsymbol{k}')E(\boldsymbol{k})}& \bar{V}_{\boldsymbol{k}'\boldsymbol{k}}^{\uparrow\downarrow\uparrow\uparrow E(\boldsymbol{k})} & \bar{V}_{\boldsymbol{k}'\boldsymbol{k}}^{\uparrow\downarrow\downarrow\downarrow E(\boldsymbol{k})} & \tilde{V}_{\boldsymbol{k}'\boldsymbol{k}}^{O(\boldsymbol{k}')E(\boldsymbol{k})} \\
    -2\bar{V}_{\boldsymbol{k}'\boldsymbol{k}}^{\uparrow\uparrow\uparrow\downarrow E(\boldsymbol{k}')} & \bar{V}_{\boldsymbol{k}'\boldsymbol{k}}^{\uparrow\uparrow\uparrow\uparrow} &   \bar{V}_{\boldsymbol{k}'\boldsymbol{k}}^{\uparrow\uparrow\downarrow\downarrow} & 2\bar{V}_{\boldsymbol{k}'\boldsymbol{k}}^{\uparrow\uparrow\uparrow\downarrow O(\boldsymbol{k}')} \\
    -2\bar{V}_{\boldsymbol{k}'\boldsymbol{k}}^{\downarrow\downarrow\uparrow\downarrow E(\boldsymbol{k}')} & \bar{V}_{\boldsymbol{k}'\boldsymbol{k}}^{\downarrow\downarrow\uparrow\uparrow} & \bar{V}_{\boldsymbol{k}'\boldsymbol{k}}^{\downarrow\downarrow\downarrow\downarrow} & 2\bar{V}_{\boldsymbol{k}'\boldsymbol{k}}^{\downarrow\downarrow\uparrow\downarrow O(\boldsymbol{k}')} \\
    \tilde{V}_{\boldsymbol{k}'\boldsymbol{k}}^{E(\boldsymbol{k}')O(\boldsymbol{k})}& \bar{V}_{\boldsymbol{k}'\boldsymbol{k}}^{\uparrow\downarrow\uparrow\uparrow O(\boldsymbol{k})} & \bar{V}_{\boldsymbol{k}'\boldsymbol{k}}^{\uparrow\downarrow\downarrow\downarrow O(\boldsymbol{k})} & \tilde{V}_{\boldsymbol{k}'\boldsymbol{k}}^{O(\boldsymbol{k}')O(\boldsymbol{k})}
    \end{pmatrix},
\end{equation}
and
\begin{align}
    \tilde{V}_{\boldsymbol{k}'\boldsymbol{k}}^{E(\boldsymbol{k}')E(\boldsymbol{k})}=& \frac{1}{4}(\tilde{V}_{-\boldsymbol{k}',-\boldsymbol{k}}+\tilde{V}_{-\boldsymbol{k}',\boldsymbol{k}}+\tilde{V}_{\boldsymbol{k}',\boldsymbol{k}}+\tilde{V}_{\boldsymbol{k}',-\boldsymbol{k}}), \\
    \tilde{V}_{\boldsymbol{k}'\boldsymbol{k}}^{E(\boldsymbol{k}')O(\boldsymbol{k})}=&  \frac{1}{4}(\tilde{V}_{-\boldsymbol{k}',-\boldsymbol{k}}-\tilde{V}_{-\boldsymbol{k}',\boldsymbol{k}}-\tilde{V}_{\boldsymbol{k}',\boldsymbol{k}}+\tilde{V}_{\boldsymbol{k}',-\boldsymbol{k}}), \\
    \tilde{V}_{\boldsymbol{k}'\boldsymbol{k}}^{O(\boldsymbol{k}')E(\boldsymbol{k})} =&  \frac{1}{4}(\tilde{V}_{-\boldsymbol{k}',-\boldsymbol{k}}+\tilde{V}_{-\boldsymbol{k}',\boldsymbol{k}}-\tilde{V}_{\boldsymbol{k}',\boldsymbol{k}}-\tilde{V}_{\boldsymbol{k}',-\boldsymbol{k}}), \\
    \tilde{V}_{\boldsymbol{k}'\boldsymbol{k}}^{O(\boldsymbol{k}')O(\boldsymbol{k})} =& \frac{1}{4}(\tilde{V}_{-\boldsymbol{k}',-\boldsymbol{k}}-\tilde{V}_{-\boldsymbol{k}',\boldsymbol{k}}+\tilde{V}_{\boldsymbol{k}',\boldsymbol{k}}-\tilde{V}_{\boldsymbol{k}',-\boldsymbol{k}}), \\
    \bar{V}_{\boldsymbol{k}'\boldsymbol{k}}^{\uparrow\downarrow\uparrow\uparrow \substack{E\\O}(\boldsymbol{k})} =& \frac{\bar{V}_{\boldsymbol{k}'\boldsymbol{k}}^{\uparrow\downarrow\uparrow\uparrow}\pm\bar{V}_{\boldsymbol{k}',-\boldsymbol{k}}^{\uparrow\downarrow\uparrow\uparrow}}{2}, \qquad
    \bar{V}_{\boldsymbol{k}'\boldsymbol{k}}^{\uparrow\downarrow\downarrow\downarrow \substack{E\\O}(\boldsymbol{k})} = \frac{\bar{V}_{\boldsymbol{k}'\boldsymbol{k}}^{\uparrow\downarrow\downarrow\downarrow} \pm \bar{V}_{\boldsymbol{k}',-\boldsymbol{k}}^{\uparrow\downarrow\downarrow\downarrow}}{2},  \\
    2\bar{V}_{\boldsymbol{k}'\boldsymbol{k}}^{\uparrow\uparrow\uparrow\downarrow \substack{E\\O}(\boldsymbol{k}')} =& \bar{V}_{\boldsymbol{k}'\boldsymbol{k}}^{\uparrow\uparrow\uparrow\downarrow} \pm \bar{V}_{-\boldsymbol{k}',\boldsymbol{k}}^{\uparrow\uparrow\uparrow\downarrow}, \qquad
    2\bar{V}_{\boldsymbol{k}'\boldsymbol{k}}^{\downarrow\downarrow\uparrow\downarrow \substack{E\\O}(\boldsymbol{k}')} = \bar{V}_{\boldsymbol{k}'\boldsymbol{k}}^{\downarrow\downarrow\uparrow\downarrow} \pm \bar{V}_{-\boldsymbol{k}',\boldsymbol{k}}^{\downarrow\downarrow\uparrow\downarrow}.
\end{align}
\end{widetext}
For temperatures close to the critical temperature, the magnitude of the gaps is negligible in $\chi_{\boldsymbol{k}\eta}$. Then, $\chi_{\boldsymbol{k}\eta} \approx \chi_{\boldsymbol{k}} = \tanh(\beta | \epsilon_{\boldsymbol{k}}|/2)/2| \epsilon_{\boldsymbol{k}}|$. Since $\chi_{\boldsymbol{k}}$ is peaked at the FS for low temperatures we expect the most important physics to occur close to the FS. In the coupling matrix $\mathcal{V}_{\boldsymbol{k}'\boldsymbol{k}}$ we restrict the momenta to lie on the FS,
and set the coupling and gap functions to nonzero values only in a small region around the FS set by the maximum magnon energy. The momentum sum is converted to integrals perpendicular and parallel to the FS. The perpendicular momentum integral is converted to an energy integral and integrated analytically while the parallel momentum integral is computed as a FS average, defined as $\langle f(\phi) \rangle_{\text{FS},\phi} = \sum_i f(\phi_i)/N_\phi$, where $N_\phi$ is the number of evenly spaced points sampled on the FS. Details are given in Refs.~\cite{AbnarMaster, Maeland2023PRL}, and the result is the gap equation in Eq.~\eqref{eq:linFSavrgap}.

\section{Higher order effects} \label{app:Higher}
In momentum space, the fermion terms originating with Eq.~\eqref{eq:Hem_i} are
\begin{align}
\label{eq:HNMrenorm}
     \frac{-2\Bar{J}S N'}{N} \sum_{r\sigma} &\bigg[ \sin\theta_r e^{-i\sigma\phi_r} \sum_{\boldsymbol{k}\in \text{EBZ},\nu}e^{-i\boldsymbol{Q}_{\nu} \cdot \boldsymbol{r}_r} c_{\boldsymbol{k}+\boldsymbol{Q}_\nu,\sigma}^\dagger c_{\boldsymbol{k},-\sigma}\nonumber\\
    &+\sigma \cos\theta_r \sum_{\boldsymbol{k}\in \text{EBZ}, \nu} e^{-i\boldsymbol{Q}_{\nu} \cdot \boldsymbol{r}_r} c_{\boldsymbol{k}+\boldsymbol{Q}_\nu,\sigma}^\dagger c_{\boldsymbol{k}\sigma} \bigg].
\end{align}
If they were included in the NM Hamiltonian, a diagonalization would be required. Then, the two degenerate bands $\epsilon_{\boldsymbol{k}}$ in the EBZ, would be folded into the MBZ, giving 30 electron bands. The lowest two bands would be approximately equal to $\epsilon_{\boldsymbol{k}}$ within the MBZ when $\Bar{J} \ll t$. Therefore, we have limited our description to FSs that fit within the MBZ, so that the terms in Eq.~\eqref{eq:HNMrenorm} only affect unoccupied states. 

Another argument for why the terms in Eq.~\eqref{eq:HNMrenorm} are negligible comes from the order in perturbation theory in which the effective electron-electron interaction has been derived. $V_{\boldsymbol{k}\boldsymbol{q}\nu\nu'}^{\alpha\alpha'\beta\beta'}/t $ is of order $(\Bar{J}/t)^2$. The bare NM band $\epsilon_{\boldsymbol{k}}$ enters in the denominator. Hence, renormalization of $\epsilon_{\boldsymbol{k}}$ to order $\Bar{J}/t$ would lead to higher order terms in the effective interaction. Assuming $\bar{J} \ll t$, their effects should be negligible. Higher order terms were also discarded in the Schrieffer-Wolff transformation \cite{SchriefferWolff, Maeland2023PRL} for the same reason.

\section{Exploring the origins of the results} \label{app:origin}
\begin{figure}
    \centering
    \includegraphics[width=\linewidth]{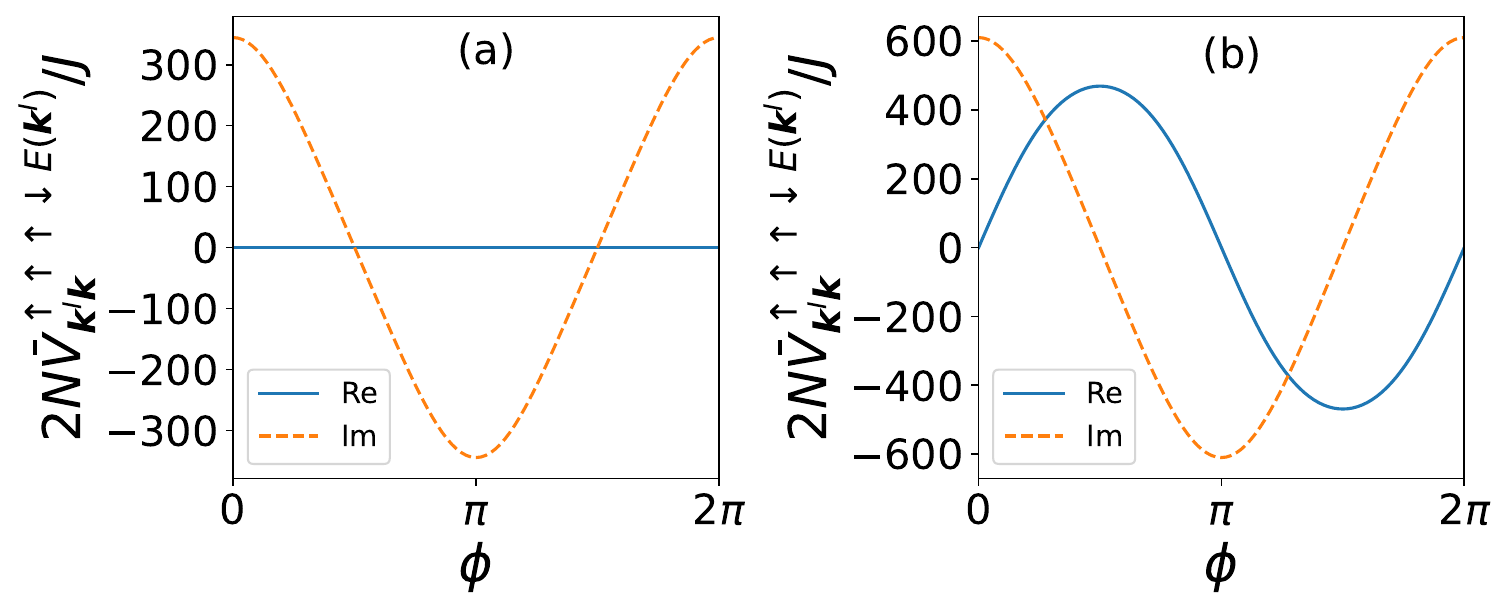}
    \caption{The negative of the 2,1 element in the coupling matrix, $-[\mathcal{V}_{\boldsymbol{k}'\boldsymbol{k}}]_{2,1} = 2\bar{V}_{\boldsymbol{k}'\boldsymbol{k}}^{\uparrow\uparrow\uparrow\downarrow E(\boldsymbol{k}')}$, is plotted as a function of the angle $\boldsymbol{k}$ makes with the $k_x$ axis as it moves around the FS, with $\boldsymbol{k}' = (k_F, 0)$ fixed. (a) 1D DMI with $D/J = 4.98$, (b) full DMI with $D/J = 2.16$. Note that the real part becomes nonzero with full DMI. The parameters are $t/J = 1000$, $\mu/t = -5.9$, $\Bar{J}/J = 50$, $K/J = 0.1$, and $S=1$. \label{fig:Vmat}}
\end{figure}

To complement the symmetry argument in the main text, we elucidate why full DMI is necessary to achieve TSC by carefully considering the details of the coupling matrix elements. 
The key difference between the 1D DMI and the full DMI is that the gaps $\Delta_{\boldsymbol{k}\sigma\sigma}$ are purely imaginary in 1D DMI, and complex in full DMI. In Fig.~\ref{fig:Vmat}, we show the 2,1 matrix element of the coupling matrix $[\mathcal{V}_{\boldsymbol{k}'\boldsymbol{k}}]_{2,1} = -2\bar{V}_{\boldsymbol{k}'\boldsymbol{k}}^{\uparrow\uparrow\uparrow\downarrow E(\boldsymbol{k}')}$ for the two DMI models. This is the matrix element that connects $\Delta_{\boldsymbol{k}\uparrow\uparrow}$ to $\Delta_{\boldsymbol{k}'\uparrow\downarrow}^{O(s)}$ in Eq.~\eqref{eq:gapeqVector}. This matrix element displays the same difference between 1D DMI and full DMI as the gap $\Delta_{\boldsymbol{k}\uparrow\uparrow}$. It is purely imaginary $p$ wave for 1D DMI, while it is complex, chiral $p$ wave for full DMI. Similar effects are found in other matrix elements as well. In the following discussion, we focus on one matrix element for clarity.

We have $ 2\bar{V}_{\boldsymbol{k}'\boldsymbol{k}}^{\uparrow\uparrow\uparrow\downarrow E(\boldsymbol{k}')} = \bar{V}_{\boldsymbol{k}'\boldsymbol{k}}^{\uparrow\uparrow\uparrow\downarrow} + \bar{V}_{-\boldsymbol{k}',\boldsymbol{k}}^{\uparrow\uparrow\uparrow\downarrow}$. 
For 1D DMI the system obeys the symmetry $C_{2y}^s \mathcal{T}$. Hence, $2\bar{V}_{\boldsymbol{k}'\boldsymbol{k}}^{\uparrow\uparrow\uparrow\downarrow E(\boldsymbol{k}')}$ should be invariant under this symmetry. In this context, $C_{2y}^s \mathcal{T}$ flips the sign of the momenta, applies a complex conjugate, but does not change the $z$ component of the spin. So, $C_{2y}^s \mathcal{T}(\bar{V}_{\boldsymbol{k}'\boldsymbol{k}}^{\uparrow\uparrow\uparrow\downarrow} + \bar{V}_{-\boldsymbol{k}',\boldsymbol{k}}^{\uparrow\uparrow\uparrow\downarrow}) = (\bar{V}_{-\boldsymbol{k}',-\boldsymbol{k}}^{\uparrow\uparrow\uparrow\downarrow*} + \bar{V}_{\boldsymbol{k}',-\boldsymbol{k}}^{\uparrow\uparrow\uparrow\downarrow*})$. We now apply the Pauli symmetry requirement $\bar{V}_{\boldsymbol{k}\boldsymbol{k}'}^{\sigma_1 \sigma_2 \sigma_3 \sigma_4} = -\bar{V}_{\boldsymbol{k},-\boldsymbol{k}'}^{\sigma_2 \sigma_1 \sigma_3 \sigma_4}$ \cite{Sigrist}, which $\bar{V}_{\boldsymbol{k}\boldsymbol{k}'}^{\sigma_1 \sigma_2 \sigma_3 \sigma_4}$ was designed to obey \cite{Maeland2023PRL}. This gives $C_{2y}^s \mathcal{T}(\bar{V}_{\boldsymbol{k}'\boldsymbol{k}}^{\uparrow\uparrow\uparrow\downarrow} + \bar{V}_{-\boldsymbol{k}',\boldsymbol{k}}^{\uparrow\uparrow\uparrow\downarrow}) = -(\bar{V}_{-\boldsymbol{k}',\boldsymbol{k}}^{\uparrow\uparrow\uparrow\downarrow} + \bar{V}_{\boldsymbol{k}'\boldsymbol{k}}^{\uparrow\uparrow\uparrow\downarrow} )^*$, which proves that $2\bar{V}_{\boldsymbol{k}'\boldsymbol{k}}^{\uparrow\uparrow\uparrow\downarrow E(\boldsymbol{k}')} = -2\bar{V}_{\boldsymbol{k}'\boldsymbol{k}}^{\uparrow\uparrow\uparrow\downarrow E(\boldsymbol{k}')*}$, i.e., it is purely imaginary for 1D DMI.

For the interested reader, we now go into details of the electron-electron coupling to see how this symmetry manifests itself in the equations.
For 1D DMI, $\Re \bar{V}_{\boldsymbol{k}'\boldsymbol{k}}^{\uparrow\uparrow\uparrow\downarrow} = - \Re \bar{V}_{-\boldsymbol{k}',\boldsymbol{k}}^{\uparrow\uparrow\uparrow\downarrow}$. Recalling the definition of the barred potentials in terms of the original potential, we see that part of the reason for the cancellation of the real part lies in the fact that $V_{\boldsymbol{k}'\boldsymbol{k}}^{\uparrow\uparrow\uparrow\downarrow } = V_{\boldsymbol{k}'\boldsymbol{k}}^{\uparrow\uparrow\downarrow\uparrow *}$. This is a nontrivial statement, and only applies to 1D DMI. 

We next recall that $V_{\boldsymbol{k}'\boldsymbol{k}}^{\uparrow\uparrow\uparrow\downarrow } = V_{\boldsymbol{k}\boldsymbol{q}\nu\bar{\nu}}^{\uparrow\downarrow\uparrow\uparrow}$ and $V_{\boldsymbol{k}'\boldsymbol{k}}^{\uparrow\uparrow\downarrow\uparrow } = V_{\boldsymbol{k}\boldsymbol{q}\nu\bar{\nu}}^{\uparrow\uparrow\uparrow\downarrow}$.
The definition of $V_{\boldsymbol{k}\boldsymbol{q}\nu\bar{\nu}}^{\alpha\alpha'\beta\beta'}$ involves factors $A_{\boldsymbol{q}n\nu\bar{\nu}}^{\alpha\alpha'\beta\beta'}$ divided by a real, spin independent denominator. Let us simplify to the FS and focus on the first term.
We have
\begin{align}
    \sum_n \frac{A_{\boldsymbol{q}n\nu\bar{\nu}}^{\uparrow\downarrow\uparrow\uparrow}}{\omega_{\boldsymbol{q}n}} &= -\frac{1}{2}\sum_{rr' n} \frac{1}{\omega_{\boldsymbol{q}n}}\Big[ g_{\nu r}^{\uparrow\downarrow} g_{\bar{\nu} r'}^{\uparrow\uparrow} U_{\boldsymbol{q},r,n}^\dagger (-V_{-\boldsymbol{q},r',n}^\dagger)\nonumber \\
    &+g_{\nu r}^{\uparrow\downarrow} g_{\nu r'}^{\uparrow\uparrow*} U_{\boldsymbol{q},r,n}^\dagger U_{\boldsymbol{q},r',n}^T  \nonumber \\
    &+g_{\bar{\nu} r}^{\downarrow\uparrow*} g_{\bar{\nu} r'}^{\uparrow\uparrow} (-V_{-\boldsymbol{q},r,n}^T) (-V_{-\boldsymbol{q},r',n}^\dagger) \nonumber \\
    &+g_{\bar{\nu} r}^{\downarrow\uparrow*} g_{\nu r'}^{\uparrow\uparrow*} (-V_{-\boldsymbol{q},r,n}^T) U_{\boldsymbol{q},r',n}^T \Big],
\end{align}
and
\begin{align}
   \sum_n \frac{A_{\boldsymbol{q}n\nu\bar{\nu}}^{\uparrow\uparrow\uparrow\downarrow}}{\omega_{\boldsymbol{q}n}} &= -\frac{1}{2}\sum_{rr' n} \frac{1}{\omega_{\boldsymbol{q}n}}\Big[ g_{\nu r}^{\uparrow\uparrow} g_{\bar{\nu} r'}^{\uparrow\downarrow} U_{\boldsymbol{q},r,n}^\dagger (-V_{-\boldsymbol{q},r',n}^\dagger)\nonumber \\
    &+g_{\nu r}^{\uparrow\uparrow} g_{\nu r'}^{\downarrow\uparrow*} U_{\boldsymbol{q},r,n}^\dagger U_{\boldsymbol{q},r',n}^T  \nonumber \\
    &+g_{\bar{\nu} r}^{\uparrow\uparrow*} g_{\bar{\nu} r'}^{\uparrow\downarrow} (-V_{-\boldsymbol{q},r,n}^T) (-V_{-\boldsymbol{q},r',n}^\dagger) \nonumber \\
    &+g_{\bar{\nu} r}^{\uparrow\uparrow*} g_{\nu r'}^{\downarrow\uparrow*} (-V_{-\boldsymbol{q},r,n}^T) U_{\boldsymbol{q},r',n}^T  \Big].
\end{align}
It is clear by looking at the GS in Fig.~\ref{fig:setupBZGS}(c) that for any sublattice $r$ there is another sublattice $r'$ such that $\phi_{r'} = (\phi_{r} + \pi) \text{~mod~} 2\pi$ and $\theta_{r'} = \pi - \theta_{r}$, i.e. the spin is flipped. With reference to the sublattice numbering in Fig.~\ref{fig:setupBZGS}(c), the sets are $(0,8),$ $(1,9)$, $(2,5),$ $(3,6),$ and $(4,7)$. In the case $(0,8)$ we define $\phi_0 = 0$, $\phi_8 = \pi$. Since these spins are up and down, $\phi_r$ is not defined, and so should not have a physical consequence. From the definition of $g_{\nu r}^{\sigma,-\sigma}$ in Eq.~\eqref{eq:gunpol}, it is clear that, e.g., $g_{\nu r'}^{\downarrow\uparrow*}$ is the complex conjugate of $g_{\nu r}^{\uparrow\downarrow}$ for these sets of $r,r'$ values. This alone does not prove that $V_{\boldsymbol{k}'\boldsymbol{k}}^{\uparrow\uparrow\uparrow\downarrow } = V_{\boldsymbol{k}'\boldsymbol{k}}^{\uparrow\uparrow\downarrow\uparrow *}$. For that, the accompanying products of factors from the transformation matrix must also be complex conjugates of each other. Numerically we find, e.g., 
\begin{equation}
\label{eq:Tinvrrp}
    \sum_n \frac{U_{\boldsymbol{q},r,n}^\dagger (-V_{-\boldsymbol{q},r',n}^\dagger)}{\omega_{\boldsymbol{q}n}} = \bqty{\sum_n  
 \frac{(-V_{-\boldsymbol{q},r,n}^T) U_{\boldsymbol{q},r',n}^T}{\omega_{\boldsymbol{q}n}}}^* 
\end{equation}
for $r,r'$ such that $\phi_{r'} = (\phi_{r} + \pi) \text{~mod~ } 2$ and $\theta_{r'} = \pi - \theta_{r}$. Then, it is clear that $V_{\boldsymbol{k}'\boldsymbol{k}}^{\uparrow\uparrow\uparrow\downarrow } = V_{\boldsymbol{k}'\boldsymbol{k}}^{\uparrow\uparrow\downarrow\uparrow *}$. 

Equation \eqref{eq:Tinvrrp} applies only to 1D DMI. With full DMI it is violated, so $V_{\boldsymbol{k}'\boldsymbol{k}}^{\uparrow\uparrow\uparrow\downarrow } \neq V_{\boldsymbol{k}'\boldsymbol{k}}^{\uparrow\uparrow\downarrow\uparrow *}$ allowing the real part of $2\bar{V}_{\boldsymbol{k}'\boldsymbol{k}}^{\uparrow\uparrow\uparrow\downarrow E(\boldsymbol{k}')}$ to become nonzero.

While the arguments for $g_{\nu r}^{\uparrow\downarrow}$ and $g_{\nu r'}^{\downarrow\uparrow*}$ being complex conjugates of each other for certain combinations of $r$ and $r'$ only rely on the GS, the transformation matrix contains information about the Hamiltonian. Full DMI results in an interaction that is complicated enough to yield TSC in the specific system we studied. By complicated enough, we mean that certain elements in the coupling matrix are complex, with phase differences between the real and imaginary part. In MI/NM heterostructures, superconductivity is induced by the magnons. So the key is that the magnons experience SOC, and that the magnetic GS is noncolinear. As discussed, 1D DMI could originate with a special linear combination of Rashba and Dresselhaus SOC. However, the symmetry $C_{2y}^s \mathcal{T}$ is preserved, which enforces trivial topology. The key feature of full DMI seems to be noncolinear DMI vectors, which ensure that symmetries of the type $C_{2}^s \mathcal{T}$ are broken. 

In MI/SC heterostructures a coplanar helical state affects the SC analogously to SOC and a magnetic field \cite{TopoSCandSkRev}, but the resulting interaction is not complicated enough to give a bulk gap. For that, noncoplanar magnetic states are needed \cite{SkTopoSCNagaosa}. For MI/NM heterostructures we believe the form of DMI is more important than the precise nature of the noncolinear magnetic ground state. Still, we stress that gap equations must be solved in any candidate system to check if the superconducting state is in fact topologically nontrivial.

\bibliography{main.bbl}

\end{document}